\documentclass[11pt]{article}
\usepackage{epsf}

\begin{document}
\def\beq{\begin{equation}}
\def\eeq{\end{equation}}
\def\bea{\begin{eqnarray}}
\def\eea{\end{eqnarray}}
\newcommand{\ket}[1]{{| #1 \rangle}}
\newcommand{\bra}[1]{{\langle #1 |}}
\newcommand{\braket}[2]{{\langle #1 | #2 \rangle}}
\newcommand{\tr}{{\rm tr}}
\newcommand{\supp}{{\rm supp}}
\newcommand{\Tr}{{\rm Tr}}
\def\thesection{\Roman{section}.}
\def\thesubsection{\Alph{subsection}.}
\renewcommand{\Im}{{\rm Im}}
\newcommand{\N}{{\rm N}}
\newcommand{\E}{{\rm E}}
\newcommand{\Id}{{\cal I}}
\newcommand{\pur}{{\rm pur}}
\newcommand{\rank}{{\rm rank}}
\newcommand{\proj}[1]{{| #1\rangle\!\langle #1 |}}
\newcommand{\ba}{\begin{array}}
\newcommand{\ea}{\end{array}}
\newtheorem{theo}{Theorem}
\newtheorem{defi}{Definition}
\newtheorem{lem}{Lemma}
\newtheorem{exam}{Example}
\newtheorem{prop}{Property}
\newtheorem{coro}{Corollary}

\title{Entanglement-Assisted Capacity of a Quantum Channel
and the Reverse Shannon Theorem}
\author{Charles H. Bennett$^*$,
Peter W. Shor$^\dag$,\\ John A. Smolin$^*$, and
Ashish V. Thapliyal$^\ddag$}
\date{}

\maketitle

\renewcommand*\thefootnote{\fnsymbol{footnote}}

\footnotetext[1]{IBM T.~J. Watson Research Center, Yorktown Heights,
NY 10598, USA; $^\dag$AT\&T Labs
-- Research, Florham Park,
NJ 07932, USA; $^\ddag$Dept.\ of Physics, U.~C.\ Santa
Barbara, Santa Barbara, CA 93106, USA.
AVT acknowledges support from US Army Research Office under grant
DAAG55-98-C0041 and DAAG55-98-1-0366. Further AVT wishes to acknowledge
support from IBM Research and David D. Awscahlom (UCSB).
CHB and JAS acknowledge support
from the National Security Agency and the Advanced Research and Development
Activity through the U.~S. Army Research Office, contract DAAG55-98-C-0041.
The material in this paper was presented in part
at the European Science Foundation Conference on
Quantum Information: Theory, Experiment, and Perspectives,
in Gdansk, Poland, July 2001.
}

\renewcommand*\thefootnote{\arabic{footnote}}

\begin{abstract} The entanglement-assisted classical capacity of a noisy
quantum channel ($C_E$) is the amount of information per channel
use that can be sent over the channel in the limit of many uses of
the channel, assuming that the sender and receiver have access to
the resource of shared quantum entanglement, which may be used up
by the communication protocol. We show that the capacity $C_E$ is
given by an expression parallel to that for the capacity of a
purely classical channel: i.e., the maximum, over channel inputs
$\rho$, of the entropy of the channel input plus the entropy of
the channel output minus their joint entropy, the latter being
defined as the entropy of an entangled purification of $\rho$
after half of it has passed through the channel. We calculate
entanglement-assisted capacities for two interesting quantum
channels, the qubit amplitude damping channel and the bosonic
channel with amplification/attenuation and Gaussian noise. We
discuss how many independent parameters are required to completely
characterize the asymptotic behavior of a general quantum channel,
alone or in the presence of ancillary resources such as prior
entanglement. In the classical analog of entanglement assisted
communication---communication over a discrete memoryless channel
(DMC) between parties who share prior random information---we show
that one parameter is sufficient, i.e., that in the presence of
prior shared random information, all DMC's of equal capacity can
simulate one another with unit asymptotic efficiency.
\end{abstract}

\section{Introduction}

The formula for the capacity of a classical channel was derived in
1948 by Shannon.  It has long been known that this formula is not
directly applicable to channels with significant quantum effects.
Extending this theorem to take quantum effects into account has
been harder than might have been anticipated; despite much recent
effort, we do not yet have a comprehensive theory for the capacity
of quantum channels. The book of Nielsen and Chuang \cite{NC} and the
survey paper \cite{Bennett-Shor} are two sources giving good
overviews of quantum information theory.  In this paper, we
advance quantum information theory by proving a capacity formula
for quantum channels which holds when the sender and receiver have
access to shared quantum entangled states which can be used in the
communication protocol.  We also present a conjecture that would
imply that, in the presence of shared entanglement, to first order
this entanglement-assisted capacity is the only quantity
determining the asymptotic behavior of a quantum channel.

A (memoryless) quantum communications channel can be viewed physically as
a process wherein a quantum system interacts with an environment
(which may be taken to initially be
in a standard state) on its way from a sender to a receiver;
it may be defined mathematically as a completely positive,
trace-preserving linear map on density operators. The theory of quantum
channels is richer and less well understood than that of classical
channels. For example, quantum channels have several distinct
capacities, depending on what one is trying to use them for, and what
additional resources are brought into play. These include

\begin{itemize}
\item The ordinary classical capacity $C$, defined as the maximum
asymptotic rate at which classical bits can be transmitted reliably
through the channel, with the help of a quantum encoder and decoder.
\item The ordinary quantum capacity $Q$, which is the maximum
asymptotic rate at which qubits can be transmitted under similar
circumstances.
\item The classically assisted quantum capacity $Q_2$, which is the
maximum asymptotic rate of reliable qubit transmission with the help
of unlimited use of a 2-way classical side channel between sender and
receiver.
\item The entanglement assisted classical capacity $C_E$, which is
the maximum asymptotic rate of reliable bit transmission with the
help of unlimited prior entanglement between the sender and receiver.
\end{itemize}

Somewhat unexpectedly, the last of these has turned out to be
the simplest to calculate, because, as we show in section II,
it is given by an expression analogous to the formula expressing
the classical capacity of a classical channel as the maximum,
over input distributions, of the input:output mutual information.
Section III calculates entanglement assisted
capacities of the amplitude damping channel and of
amplifying and attenuating bosonic channels with Gaussian noise.

We return now to a general discussion of quantum channels and
capacities, in order to provide motivation for section IV of
the paper, on what we call the reverse Shannon theorem.

Aside from the constraints $Q\!\leq\!C\!\leq\!C_E$, and $Q\!\leq\!Q_2$,
which are obvious consequences of the definitions, the four capacities
appear to vary rather independently.  It is conjectured that
$Q_2 \! \leq \! C$, but this has not been proved to date.  Except in
special cases, it is not possible, without knowing the parameters of a
channel, to infer any one of its four capacities from the other three. This
independence is illustrated in Table~\ref{chan-table}, which compares the
capacities of several simple channels for which they are known exactly.
The channels incidentally illustrate four different degrees of qualitative
quantumness: the first can carry qubits unassisted, the second
requires classical assistance to do so, the third has no quantum
capacity at all but still exhibits quantum behavior in that its
capacity is increased by entanglement, while the fourth is completely
classical, and so unaffected by entanglement.

\begin{table}

\label{chan-table}

\caption{Capacities of several quantum channels.}

\begin{center}
{\small
\begin{tabular}{l|c c c c}
Channel &$Q$ & $Q_2$ & $C$ & $C_E$ \\ \hline
Noiseless qubit channel         &   $\;$1$\;$  & $\;$1$\;$ & 1       & 2     \\
50\% erasure qubit channel     & 0  & 1/2 & 1/2     & 1      \\
2/3 depolarizing qubit channel  & 0  & 0   &  0.0817$^*$ & 0.2075 \\
Noiseless bit channel =  & 0  & 0   & 1       & 1      \\
100\% dephasing qubit channel & &  & & \\
\hline
\end{tabular}
}
\parbox{3.5in}{\footnotesize
$^*$Proved in \protect{\cite{CKing}}.}
\end{center}
\end{table}

Contrary to an earlier conjecture of ours, we have found channels for
which $Q\!>\! 0$
but $C \!=\! C_E$.  One example is a channel mapping three qubits to
two qubits which is switched between two different behaviors
by the first input qubit.  The channel operates as follows:
The first qubit is measured
in the $\ket{0},\, \ket{1}$ basis.  If the result is $\ket{0}$, then the
other two qubits are dephased (i.e., measured in the $\ket{0}$, $\ket{1}$
basis) and transmitted as classical bits; if the
result is $\ket{1}$, the first qubit is transmitted intact and the
second qubit is replaced by the completely mixed state.
This channel has $Q \!= \!Q_2 \!= \!1$ (achieved by setting the first
qubit to $\ket{1}$) and
$C \!= \!C_E \!=\! 2$.

This complex situation naturally raises the question of how many
independent parameters are needed to characterize the important
asymptotic, capacity-like properties of a general quantum channel. A
full understanding of quantum channels would enable us to calculate not
only their capacities, but more generally, for any two channels ${\cal
M}$ and ${\cal N}$, the asymptotic efficiency (possibly zero) with which
${\cal M}$ can simulate ${\cal N}$, both alone and in the presence of
ancillary resources such as classical communication or shared
entanglement.

One motivation for studying communication in the presence of ancillary
resources is that it can simplify the classification of channels'
capacities to simulate one another.  This is so because if a
simulation is possible without the ancillary resource, then the
simulation remains possible with it, though not necessarily vice versa.
For example, $Q$ and $C$ represent a channel's asymptotic efficiencies
of simulating, respectively, a noiseless qubit channel and a noiseless
classical bit channel. In the absence of ancillary resources these two
capacities can vary independently, subject to the constraint $Q\leq C$,
but in the presence of unlimited prior shared entanglement, the relation
between them becomes fixed: $C_E=2Q_E$, because shared entanglement
allows a noiseless 2-bit classical channel to simulate a noiseless
1-qubit channel and vice versa (via teleportation \cite{teleport} and
superdense coding \cite{superdense}).

We conjecture that prior entanglement so simplifies the complex
landscape of quantum channels that only a single free parameter remains.
Specifically, we conjecture that in the presence of unlimited prior
entanglement,
any two quantum channels of equal $C_E$ could simulate one another with
unit asymptotic efficiency. Section IV proves a classical analog of this
conjecture, namely that in the presence of prior random information
shared between sender and receiver, any two discrete memoryless
classical channels (DMC's) of equal capacity
can simulate one another with unit asymptotic
efficiency. We call this the classical reverse Shannon theorem because
it establishes the ability of a noiseless classical DMC to simulate
noisy ones of equal capacity, whereas the ordinary Shannon theorem
establishes that noisy DMC's can simulate noiseless ones of equal
capacity.

Another ancillary resource---classical communication---also simplifies
the landscape of quantum channels, but probably not so much. The
presence of unlimited classical communication does allow certain
otherwise inequivalent pairs of channels to simulate one another (for
example, a noiseless qubit channel and a 50\% erasure channel on
4-dimensional Hilbert space), but it does not render all
channels of equal $Q_2$ asymptotically equivalent. So-called
bound-entangled channels~\cite{boundecH,boundecUS} have $Q_2\!=\!0$, but unlike
classical channels (which also have $Q_2\!=\!0$) they can be
used to prepare bound entangled states, which are entangled but cannot
be used to prepare any pure entangled states.  Because the distinction
between bound entangled and unentangled states does not vanish
asymptotically, even in the presence of unlimited classical
communication \cite{VC}, bound-entangled and classical channels must
be asymptotically inequivalent, despite having the same $Q_2$.

The various capacities of a quantum channel ${\cal N}$ may be
defined within a common framework, \beq C_X({\cal
N})=\lim_{\epsilon\rightarrow0}\limsup_{n\rightarrow\infty}\;
\{\frac{m}{n}: \exists_{{\cal A}}\exists_{{\cal B}}
\forall_{\psi\in \Gamma_m}\; F(\psi,{\cal A},{\cal B},{\cal
N})\;>1\!-\!\epsilon\;\;\}. \label{framework} \eeq Here $C_X$ is a
generalized capacity; ${\cal A}$ is an encoding subprotocol, to be
performed by Alice, which receives an $m$-qubit state $\psi$
belonging to some set $\Gamma_m$ of allowable inputs to the entire
protocol, and produces $n$ possibly entangled inputs to the
channel ${\cal N}$; ${\cal B}$ is a decoding subprotocol, to be
performed by Bob, which receives $n$ (possibly entangled) channel
outputs and produces an $m$-qubit output for the entire protocol;
finally $F(\psi,{\cal A},{\cal B},{\cal N})$ is the {\em fidelity}
of this output relative to the input $\psi$, i.e., the probability
that the output state would pass a test determining whether it is
equal to the input (more generally, the fidelity of one mixed
state $\rho$ relative to another $\sigma$ is
$F=(\tr(\sqrt\rho\;\sigma\sqrt{\rho}))^2$). Different capacities
are defined depending on the specification of $\Gamma$, ${\cal A}$
and ${\cal B}$. The classical capacities $C$ and $C_E$ are defined
by restricting $\psi$ to a standard orthonormal set of states,
without loss of generality the ``Boolean'' states labelled by bit
strings $\Gamma_m=\{\ket{0},\ket{1}\}^{\otimes m}$; for the
quantum capacities $Q$ and $Q_2$, $\Gamma_m$ is the entire $2^m$
dimensional Hilbert space ${\cal H}_2^{\otimes m}$. For the simple
capacities $Q$ and $C$, the Alice and Bob subprotocols are
completely-positive trace-preserving maps from ${\cal
H}_2^{\otimes m}$ to the input space of ${\cal N}^{\otimes n}$,
and from the output space of ${\cal N}^{\otimes n}$ back to ${\cal
H}_2^{\otimes m}$. For $C_E$ and $Q_2$, the subprotocols are more
complicated, in the first case drawing on a supply of ebits
(maximally entangled pairs of qubits) shared beforehand between
Alice and Bob, and in the latter case making use of a 2-way
classical channel between Alice and Bob. The definition of $Q_2$
thus includes interactive protocols, in which the $n$ channel uses
do not take place all at once, but may be interspersed with rounds
of classical communication.

The classical capacity of a classical discrete memoryless channel
is also given by an expression of the same form, with $\psi$
restricted to Boolean values; the encoder ${\cal A}$, decoder
${\cal B}$, and channel ${\cal N}$ all being restricted to be
classical stochastic maps; and the fidelity $F$ being defined as
the probability that the (Boolean) output of ${\cal B}({\cal
N}^{\otimes n}({\cal A}(\psi)))$ is equal to the input $\psi$. We
will sometimes indicate these restrictions implicitly by using
upper case italic letters (e.g.\ $N$) for classical stochastic maps,
and lower case italic letters (e.g. $x$) for classical discrete
data. The definition of classical capacity would then be \beq
C(N)=\lim_{\epsilon\rightarrow0}\limsup_{n\rightarrow\infty}\;
\{\frac{m}{n}: \exists_{A}\exists_{B}\; \forall_{x\in \{0,1\}^m}
F(x,A,B,N)\;
>1\!-\!\epsilon\;\;\}.
\eeq A classical stochastic map, or classical channel, may be
defined in quantum terms as one that is completely dephasing in
the Boolean basis both with regard to its inputs and its outputs.
A channel, in other words, is classical if and only if it can be
represented as a composition \beq N={\cal D'GD}
\label{classchan}\eeq of the completely dephasing channel ${\cal
D}$ on the input Hilbert space, followed by a general quantum
channel ${\cal G}$, followed by the completely dephasing channel
${\cal D'}$ on the output Hilbert space (a completely dephasing
channel is one that makes a von Neumann measurement in the Boolean
basis and resends the result of the measurement). Dephasing only
the inputs, or only the outputs, is in general insufficient to
abolish all quantum properties of a quantum channel ${\cal G}$.

The notion of capacity may be further generalized to define a
capacity of one channel
${\cal N}$ to simulate another channel  ${\cal M}$.  This may be
defined as

\beq C_X({\cal N,M}) = \lim_{\epsilon \rightarrow 0}
\limsup_{n\rightarrow \infty} \{\frac{m}{n}: \exists_{\cal A,B}
\forall_{\psi\in H_M^{\otimes m}}\; F({\cal M}^{\otimes
m}(\psi),{\cal A},{\cal B},{\cal N})\;>1\!-\!\epsilon\;\;\},
\label{MNsim}\eeq where ${\cal A}$  and ${\cal B}$ are
respectively Alice's and Bob's subprotocols which together enable
Alice to receive an input $\psi$ in $H_M^{\otimes m}$ (the tensor
product of $m$ copies of the input Hilbert space $H_M$ of the
channel ${\cal M}$ to be simulated) and, making $n$ forward uses
of the simulating channel ${\cal N}$, allow Bob to produce some
output state, and $F({\cal M}^{\otimes m}(\psi),{\cal A},{\cal
B},{\cal N})$ is the fidelity of this output state with respect to
the state that would have been generated by sending the input
$\psi$ through ${\cal M}^{\otimes m}$.

These definitions of capacity are all asymptotic, depending on the
properties of ${\cal N}^{\otimes n}$ in the limit
$n\!\rightarrow\!\infty$. However, several of the capacities are given
by, or closely related to, non-asymptotic expressions involving input
and output entropies for a single use of the channel. Figure 1 shows a
scenario in which a quantum system $Q$, initially in mixed state
$\rho$, is sent through the channel, emerging in a mixed state ${\cal
N}(\rho)$. It is useful to think of the initial mixed state as being
part of an entangled pure state $\Phi_\rho^{QR}$ where $R$ is some
reference system that is never operated upon physically. Similarly the
channel can be thought of as a unitary interaction $U$ between the
quantum system $Q$ and some environment subsystem $E$, which is
initially supplied in a standard pure state $0^E$, and leaves the
interaction in a mixed state ${\cal E}(\rho)^E$. Thus ${\cal N}$ and
${\cal E}$ are completely positive maps relating the final states of
the channel output and environment, respectively, to the initial state
of the channel input, when the initial state of the environment is held
fixed. The mnemonic superscripts $Q,R,E$ indicate, when necessary, to
what system a density operator refers.

\begin{figure}[tbp]
\epsfxsize=4in\hspace*{1in} \epsfbox{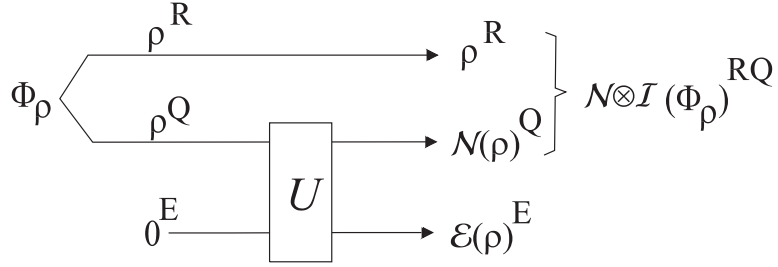}\medskip
\caption{A quantum system Q in mixed state $\rho$ is sent through
the noisy channel ${\cal N}$, which may be viewed as a unitary
interaction $U$ with an environment E. Meanwhile a purifying
reference system R is sent through the identity channel ${\cal
I}$. The final joint state of RQ has the same entropy as the final
state ${\cal E}(\rho)$ of the environment.} \label{bigcef1}
\end{figure}

Under these circumstances
three useful von Neumann entropies may be defined, the input entropy
\[
H(\rho^Q)=-\tr\rho^Q\log_2\rho^Q,
\]
the output entropy
\[
H({\cal N}(\rho)^Q),
\]
and the {\em entropy exchange}
\[
H(({\cal N}\otimes{\cal I})\Phi_\rho^{QR})=H({\cal E}(\rho)^E).
\]

The complicated left side of the last equation represents the entropy of
the joint state of the subsystem $Q$ which has been through the channel,
and the reference system $R$, which has not, but may still be more or
less entangled with it. The density operator $({\cal N}\otimes{\cal
I})\Phi_\rho$ is the quantum analog of a joint input:output probability
distribution, because it has ${\cal N}(\rho)$ and $\rho$ as its partial
traces. Without the reference system, the notion of a joint input:output
mixed state would be problematic, because the input and output are not
present at the same time, and the no-cloning theorem prevents Alice from
retaining a spare copy of the input to be compared with the one sent
through the channel. The entropy exchange is also equal to the final
entropy of the environment $H({\cal E}(\rho))$, because the tripartite
system $QRE$ remains throughout in a pure state; making its two
complementary subsystems $E$ and $QR$ always isospectral. The relations
between these entropies and quantum channels have been well reviewed by
Schumacher and Nielsen ~\cite{SN} and by Holevo and Werner~\cite{HW99}.

By Shannon's theorem, the capacity of a classical channel $N$ is the
maximum, over input distributions, of the input:output mutual
information, in other words the input entropy plus the output
entropy less the joint entropy of input and output. The quantum
generalization of mutual information for a bipartite mixed state
$\rho^{AB}$, which reduces to classical mutual information when
$\rho^{AB}$ is diagonal in a product basis of the two subsystems, is
\[
H(\rho^A)+H(\rho^B)-H(\rho^{AB}).
\]
where
\[
\rho^A \! =\tr_B\rho^{AB}\qquad\hbox{and}\qquad
\rho^B \! =\tr_A\rho^{AB}.
\]
In terms of Figure \ref{bigcef1}, the classical capacity of a
classical channel (cf.~eq. (\ref{classchan})) can be expressed as
\beq C(N) =
\max_{\rho\in\Delta} H(\rho) + H(N(\rho)) - H((N\otimes{\cal
I})(\Phi_\rho)) \eeq
where $\Delta$ is the class of density
operators on the channel's input Hilbert space that are diagonal
in the Boolean basis. The third term (entropy exchange), for a
classical channel $N$, is just the joint Shannon entropy of the
classically correlated Boolean input and output, because the von
Neumann entropies reduce to Shannon entropies when evaluated in
the Schmidt basis of $\Phi_\rho$, with respect to which all states
are diagonal. The restriction to classical inputs $\rho\in\Delta$
can be removed, because any non-diagonal elements in $\rho$ would
only reduce the first term, while leaving the other two terms
unchanged, by virtue of the diagonality-enforcing properties of
the channel.

Thus, the expression
\begin{equation}
\max_{\rho\in{\cal H}_{\rm{in}}} H(\rho) + H({\cal N}(\rho))
-H(({\cal N}\otimes{\cal I})\Phi_\rho),
\label{CE-capacity}
\end{equation}
is a natural generalization to quantum channels ${\cal N}$ of a
classical channel's maximal input:output mutual information, and
it is equal to the classical capacity whenever ${\cal N}$ is classical,
as defined previously in this section.

One might hope that this expression continues to give the classical
capacity of a general quantum channel ${\cal N}$, but that is not so,
as can be seen by considering the simple case ${\cal N=I}$ of a
noiseless qubit channel.  Here the maximum is attained on a uniform
input mixed state $\rho=I/2$, causing the first two terms each
to have the value 1 bit, while the last term is zero, giving a
total of 2 bits. This is not the ordinary classical capacity of the
noiseless qubit channel, which is equal to 1 bit,
but rather its entanglement-assisted capacity $C_E({\cal
N})$.  In the next section we show that this is true of quantum channels
in general, as stated by the following theorem.

\begin{theo}
Given a quantum channel ${\cal N}$, then the entanglement-assisted capacity
of the quantum channel $C_E$ is equal to the
maximal quantum mutual information
\begin{equation}
C_E = \max_{\rho\in{\cal H}_{\rm{in}}} H(\rho) + H({\cal N}(\rho))
-H(({\cal N}\otimes{\cal I})\Phi_\rho).
\label{CE-capacity-theorem}
\end{equation}

Here the capacity $C_E$ is defined as the supremum of
Eq.~(\ref{framework}) when $\psi$ ranges over Boolean states and
${\cal A}$, ${\cal B}$ over all protocols where Alice and Bob
start with an arbitrarily large number of shared EPR
pairs\footnote{It is sufficient to use standard EPR
pairs---maximally entangled two-qubit states---as the entanglement
resource because any other entangled state can be efficiently
prepared from EPR pairs by the process of entanglement dilution
using an asymptotically negligible $o(n)$ amount of forward
classical communication~\cite{LP99}.},
but have no access to any communication channels other than ${\cal
N}$.
\end{theo}

Another capacity theorem which has been proven for quantum channels is the
Holevo-Schumacher-Westmoreland theorem \cite{Holevo,SW}, which says that
if the signals that Bob receives are constrained to lie in a set
of quantum states $\rho'_i$, where Alice chooses $i$ (for example, by
supplying input state $\rho_i$ to the channel ${\cal N}$)
then the capacity is given by
\begin{equation}
C_H(\{\rho'_i\})=
 H(\sum_i p_i\rho'_i) - \sum_i p_i H(\rho'_i).
\label{Holevo}
\end{equation}
This gives a means to calculate a constrained classical capacity
for a quantum channel ${\cal N}$
if the sender is not allowed to use entangled inputs: the channel's
Holevo capacity $C_H({\cal N})$ being defined as the maximum of
$C_H(\{{\cal N}(\rho_i)\})$ over all
possible sets of input states $\{\rho_i\}$.  We will be using this theorem
extensively in the proof of our entanglement-assisted capacity bound.

In our original paper \cite{CEprl}, we proved the formula (\ref{CE-capacity})
for certain special cases, including the depolarizing channel and the
erasure channel.
We did this by sandwiching the entanglement-assisted capacity
between two other capacities, which for certain channels turned out to
be equal.  The higher of these two capacities we called the
forward classical communication cost via teleportation,
($FCCC_{Tp}$), which is the amount of
forward classical communication needed to simulate the channel ${\cal N}$
by teleporting over a noisy classical channel.  The lower of these
two bounds we called $C_{Sd}$, which is the capacity obtained by
using the noisy quantum channel ${\cal N}$ in the superdense coding
protocol.  We have that $C_{Sd} \leq C_E \leq FCCC_{Tp}$.
Thus, if $C_{Sd} = FCCC_{Tp}$ for a channel, we have obtained
the entanglement-assisted capacity of the channel.  In order
for this argument to work, we needed the classical reverse Shannon
theorem, which says that a noisy classical
channel can be simulated by a noiseless classical channel of the same
capacity, as long as the sender and receiver have access to shared random
bits.  We needed this theorem because the causality argument showing
that EPR pairs do not add to the capacity of a classical channel
appears to work only for noiseless channels.  We sketched the proof of the
classical reverse Shannon theorem in
our previous paper, and give it in full in this paper.

In our previous paper, the bounds $C_{Sd}$ and $FCCC_{Tp}$ are both
computed using single-symbol protocols; that is, both the superdense coding
protocol and the simulation of the channel by teleportation via a noisy
classical channel are carried out with a single use of the channel.
The capacity is then obtained using the classical
Shannon formula for a classical channel associated with these protocols.
In this paper, we obtain bounds using multiple-symbol protocols, which
perform entangled operations on many uses of the channel.  We then perform
the capacity computations using the Holevo-Schumacher-Westmoreland
formula (\ref{Holevo}).

\section{Formula for Entanglement Assisted Classical Capacity}

Assume we have a quantum channel
${\cal N}$ which maps a Hilbert space ${\cal H}_{\rm{in}}$
to another Hilbert space ${\cal H}_{\rm{out}}$.  Let $C_E$ be the classical
capacity of the channel when the sender and receiver have an unbounded
supply of EPR pairs
to use in the communication protocol.  This section proves
that the entanglement-assisted capacity of a channel is the maximum
quantum mutual information attainable between the two parts of an entangled
quantum state, one part of which has been passed through the channel.
That is,
\begin{equation}
C_E({\cal N}) = \max_{\rho \in {\cal H}_{\rm{in}}} \ \
 H(\rho) + H({\cal N}(\rho)) - H(({\cal N} \otimes {\cal I})\Phi_\rho),
\label{CE}
\end{equation}
where $H(\rho)$ denotes the von Neumann entropy of a density matrix
$\rho \in {\cal H}_{\mathrm{in}}$,
$H({\cal N}(\rho))$ denotes the von Neumann entropy of
the output when $\rho$ is input into the channel, and
$H(({\cal N} \otimes {\cal I})\Phi_\rho)$ denotes the von Neumann
entropy of a purification $\Phi_\rho$ of $\rho$ over a reference
system ${\cal H}_{\mathrm{ref}}$, half of which (${\cal H}_{\mathrm{in}}$)
has been sent through
the channel ${\cal N}$ while the other half (${\cal H}_{\mathrm{ref}}$)
has been sent through the
identity channel ${\cal I}$ (this corresponds to the portion of the
entangled state that Bob holds at the start of the protocol).  Here, we have
$\Phi_\rho \in {\cal H}_{\rm{in}} \otimes {\cal H}_{\rm{ref}}$
and $\Tr_{\rm{ref}} \Phi_\rho = \rho$.
All purifications of $\rho$ give the same entropy in this
formula\footnote{This is a consequence of the fact that
any two purifications of a given density matrix can be mapped to each other
by a unitary transformation of the reference system \cite{HJW}.},
so we need not specify which one we use.  As pointed out earlier,
the right hand side of
Eq.~(\ref{CE}) parallels the expression for capacity of a classical
channel as the maximum, over input distributions, of the input:output
mutual information.

Lindblad~\cite{Lindblad91}, Barnum et al.~\cite{BNS97}, and Adami and
Cerf~\cite{CandA} characterized
several important properties of the quantum mutual information,
including positivity, additivity and the data processing inequality.
Adami and Cerf argued that the
right side of Eq.~(\ref{CE}) represents an important channel property,
calling it the channel's ``von Neumann capacity'', but they did not
indicate what kind of communication task this capacity represented the
channel's asymptotic efficiency for doing. Now we know that
it is the channel's efficiency for transmitting classical information
when the sender and receiver share prior entanglement.

In our demonstration that Eq.~(\ref{CE}) is indeed the correct
expression for entanglement assisted classical capacity, the first
subsection gives an entanglement assisted classical communication
protocol which can asymptotically achieve the
rate ${\mathrm{RHS}} - \epsilon$ for any $\epsilon$.
The second subsection gives a proof of a crucial lemma
on typical subspaces needed in the
first subsection.  The third subsection shows that the right hand
side of Eq.~(\ref{CE}) is indeed an upper bound for $C_E({\cal N})$.
The fourth subsection proves several entropy inequalities
that are used in the third subsection.

\subsection{Proof of the Lower Bound}
In this section, we will prove the inequality
\beq
C_E({\cal N}) \geq \max_{\rho \in {\cal H }_{\rm{in}}} \ \
 H(\rho) + H({\cal N}(\rho)) - H({\cal N}\!\!\otimes\!{\cal I}\; (\Phi_\rho)).
\label{lower-bound}
\eeq
We first show the inequality
\beq
C_E({\cal N}) \geq
 H(\rho) + H({\cal N}(\rho)) - H({\cal N}\!\!\otimes\!{\cal I}\; (\Phi_\rho))
\label{nomax}
\eeq
for the special case where
$\rho = \frac{1}{d} I$, where $d = \dim {\cal H}_{\rm{in}} $,
$I$ is the identity matrix, and $\Phi_\rho$ is a maximally entangled
state.  We then use this special case to show that the inequality
(\ref{nomax}) still holds when $\rho$ is any projection matrix.  We
finally use the case where $\rho$ is a projection matrix to
prove the inequality in the general case of arbitrary $\rho$, showing
(\ref{lower-bound}); we do this by taking $\rho'$
to be the projection onto the typical subspace of $\rho^{\otimes n}$, and
using $\rho'$ and ${\cal N}^{\otimes n}$ in the inequality (\ref{nomax}).

The coding protocol we use for the special case given above, where
$\rho = \frac{1}{d}I$, is essentially the same as the protocol used for
quantum superdense coding \cite{superdense},
which procedure yields the entanglement-assisted
capacity in the case of a noiseless quantum
channel.  The proof that the formula (\ref{nomax}) holds for $\rho=I/d$,
however, is quite different from and somewhat more complicated than the
proof that superdense coding works.  Our proof uses Holevo's formula
(\ref{Holevo}) for
quantum capacity to compute the capacity achieved by our protocol.
This protocol is the same as that given in our earlier paper
on $C_E$ \cite{CEprl}, although our proof is different; the earlier proof
only applied to certain quantum channels, such as those that commute
with teleportation.

We need to use the generalization
of the Pauli matrices to
$d$ dimensions.  These are the matrices used in the $d$-dimensional
quantum teleportation scheme \cite{teleport}.  There are $d^2$ of these
matrices, which
are given by $U_{j,k} = T^j R^k$, for the matrices $T$ and $R$ defined
by their entries as
\beq
T_{a,b} = \delta_{a,\,b-1 {\rm{\,mod\,}}d} {\rm{\ \ and\ \ }}
R_{a,b} =  e^{2 \pi i a/d} \delta_{a,b}
\eeq
as in \cite{AK}.
To achieve the capacity given by the above formula (\ref{nomax})
with $\rho = I/d$, Alice and Bob start
by sharing
a $d$-dimensional maximally entangled state $\phi$.  Alice applies one of the
$d^2$ transformations $U_{j,k}$ to her part of $\phi$, and then sends
it through the channel ${\cal N}$.
Bob gets one of the $d^2$ quantum
states $({\cal N}\otimes {\cal I}) (U_{j,k} \otimes {\cal I}) \phi$.  It is straightforward to
show that averaging over the matrices $U_{j,k}$ effectively
disentangles Alice's and Bob's pieces, so we obtain
\begin{eqnarray}
\sum_{j,k=1}^d ({\cal N}\otimes {\cal I}) (U_{j,k} \otimes {\cal I}) \phi
&=& {\cal N}(\Tr_B \phi) \otimes \Tr_A \phi \nonumber \\
&=& {\cal N}(\rho) \otimes \rho
\end{eqnarray}
where $\rho = \frac{1}{d}I$.  The entropy of this quantity is the first
term of Holevo's formula \ref{Holevo}, and
gives the first
two terms of (\ref{nomax}).  The entropy of each of the
$d^2$ states $({\cal N}\otimes {\cal I})(U_{j,k} \otimes {\cal I})\phi$ is
$H(({\cal N} \otimes {\cal I})(\Phi_\rho))$, since each of the
$(U_{j,k}\otimes {\cal I})(\phi)$
is a purification of $\rho$.  This entropy is the second term of
Holevo's formula \ref{Holevo}, and gives the third term of (\ref{nomax}).
We thus obtain the formula when $\rho = \frac{1}{d}I$.

The next step is to note that the inequality~(\ref{nomax})
also holds if
the density matrix $\rho$ is a projection onto any subspace of
${\cal H}_{\rm{in}}$.
The proof is exactly the same as for $\rho = \frac{1}{d}I$.
In fact, one can
prove this case by using the above result.
By restricting ${\cal H}_{\rm{in}}$ to the support of $\rho$,
which we can denote by ${\cal H}'$, and by
restricting ${\cal N}$ to act only on ${\cal H}'$, we obtain
a channel ${\cal N}'$ for which $\rho' = \frac{1}{d_{\mathrm{in}}'} I$.

We now must show that (\ref{nomax}) holds for arbitrary $\rho$.
This is the most difficult part of the proof.
For this step we need a little more notation.
Recall that we can assume that any quantum map ${\cal N}$ can be implemented
via a unitary transformation
${\cal U}$ acting on the system ${\cal H}_{\rm{in}}$ and some environment
system ${\cal H}_{\rm{env}}$, where ${\cal H}_{\rm{env}}$ starts in some
fixed initial state.
We introduce ${\cal E}$, which is the completely positive map taking
${\cal H}_{\rm{in}}$ to
${\cal H}_{\rm{env}}$ by first
applying ${\cal U}$ and tracing out everything but ${\cal H}_{\rm{env}}$.
We then have
\beq
H({\cal E}(\rho)) = H( ({\cal N}\otimes {\cal I}) \Phi_\rho )
\label{entropy-exchange}
\eeq
where $\rho$ is
a density matrix over ${\cal H}_{\rm{in}}$ and $\Phi_\rho$ is a
purification of $\rho$.  Recall (from footnote 2) that
this does not depend on which
purification $\Phi_\rho$ of $\rho$ is used.

As our argument involves
typical subspaces, we first give some facts about typical subspaces.
For technical reasons,\footnote{Our proof of
Lemma~\ref{typical-lemma} does not appear to
work for entropy-typical subspaces unless these subspaces are modified
by imposing a somewhat unnatural-looking extra condition.  This will be
discussed later.} we use frequency-typical
subspaces.  For any $\epsilon$ and $\delta$ there is a large enough $n$ such
the Hilbert space ${\cal H}^{\otimes n}$ contains a typical subspace $T$
(which is the span of typical eigenvectors of $\rho$)
such that
\begin{enumerate}
\item
$\Tr \, \Pi_T\,  \rho^{\otimes n}\,  \Pi_T > 1-\epsilon$,
\item
The eigenvalues $\lambda$ of $\Pi_T\,  \rho^{\otimes n}\, \Pi_T$ satisfy
\[
2^{-n (H(\rho)+\delta)} \leq \lambda \leq
2^{-n (H(\rho)-\delta)}\ ,
\]
\item
$(1-\epsilon) 2^{n (H(\rho) - \delta )} \leq
\dim T \leq 2^{n (H(\rho) + \delta)}$.
\end{enumerate}
\medskip

Let $T_n \subset {\cal H}^{\otimes n}$ be the typical subspace
corresponding to $\rho^{\otimes n}$, and let $\pi_{T_n}$ be the
normalized density matrix proportional to the projection onto $T_n$.
It follows from well-known facts about typical subspaces that
\[
\lim_{n \rightarrow \infty} \frac{1}{n} H(\pi_{T_n}) = H(\rho).
\]
We can also show the following lemma.  We delay giving the proof of
this lemma until after the proof of the theorem.
\begin{lem}
\label{typical-lemma}
Let ${\cal N}$ be a noisy quantum channel and $\rho$ a density matrix
on the input space of this channel.  Then we can find a sequence
of frequency typical subspaces $T_n$ corresponding to $\rho^{\otimes n}$,
such that if $\pi_{T_n}$ is the unit trace density matrix proportional
to the projection onto $T_n$, then
\beq
\lim_{n \rightarrow \infty} \frac{1}{n} H({\cal N}^{\otimes n}(\pi_{T_n})) = H({\cal N}(\rho)).
\eeq
\end{lem}

Applying the lemma to the map onto the environment similarly gives
\beq
\lim_{n \rightarrow \infty} \frac{1}{n} H({\cal E}^{\otimes n}(\pi_{T_n})) = H({\cal E}(\rho)).
\eeq
Thus,
if we consider the quantity
\beq
\frac{1}{n}\left[ H(\pi_{T_n}) +  H({\cal N}^{\otimes n}(\pi_{T_n}))
- H({\cal E}^{\otimes n}(\pi_{T_n}))\right]
\eeq
we see that it converges to
\beq
H(\rho) + H({\cal N}(\rho)) - H({\cal E}(\rho)),
\eeq
which the identity (\ref{entropy-exchange}) shows is equal to the
desired quantity (\ref{CE}).  This concludes the proof of the
lower bound.

One more matter to be cleared up is the form of the prior entanglement
to be shared by Alice and Bob. The most standard form of entanglement is
maximally entangled pairs of qubits (``ebits''), and it is natural
to use them as the entanglement resource in defining $C_E$. However,
Eq.~(\ref{CE}) involves the entangled state $\Phi_\rho$, which is typically
not a product of ebits. This is no problem, because, as Lo
and Popescu~\cite{LP99} showed, many copies of two
entangled pure states having an equal entropy of entanglement can
be interconverted not only with unit
asymptotic efficiency, but in a way that requires an asymptotically
negligible amount of (one-way) classical communication, compared to the
amount of entanglement processed.  Thus the definition of $C_E$ is
independent of the form of the entanglement resource, so long as it is a
pure state.  As it turns out, the lower bound proof does not
actually require construction of $\Phi_\rho$ itself, but merely a
sequence of maximally entangled states on high-dimensional typical
subspaces $T_n$ of tensor powers of $\Phi_\rho$.  These maximally
entangled states can be prepared from standard ebits with
arbitrarily high fidelity and no classical communication~\cite{BBPS}.

\subsection{Proof of Lemma \ref{typical-lemma}}

In this section, we prove
\setcounter{lem}{0}
\begin{lem}
Suppose $\rho$ is a density matrix over a Hilbert space ${\cal H}$
of dimension $d$, and $\cal N$,
$\cal E$, are two trace-preserving completely positive maps.
Then there is a sequence of frequency-typical subspaces
$T_n \subset {\cal H}^{\otimes n}$ corresponding to $\rho^{\otimes n}$
such that
\beq
\lim_{n \rightarrow \infty} \frac{1}{n} \dim T_n = H(\rho),
\label{Slimit}
\eeq
\beq
\lim_{n \rightarrow \infty} \frac{1}{n} H({\cal N}^{\otimes n}(\pi_{T_n}))
=  H({\cal N}(\rho)),
\label{nlimit}
\eeq
and
\beq
\lim_{n \rightarrow \infty} \frac{1}{n} H({\cal E}^{\otimes n}(\pi_{T_n}))
=  H({\cal E}(\rho)),
\label{elimit}
\eeq
where $\pi_{T_n}$ is the projection matrix onto $T_n$ normalized to
have trace 1.
\end{lem}

For simplicity, we will prove this lemma with only the conditions (\ref{Slimit})
and (\ref{nlimit}).  Altering the proof to also
obtain the condition (\ref{elimit})
is straightforward, as we treat the map ${\cal E}$ in exactly the same
manner as the map ${\cal N}$, and need only make sure that both formulas
(\ref{nlimit}) and (\ref{elimit}) converge.

Our proof is based on several previous
results in quantum information theory.  For the proof of the $\leq$ direction
in Eq.~(\ref{nlimit}),
we show that a source producing states with average
density matrix ${\cal N}^{\otimes n}(\pi_{T_n})$
can be compressed into $n H({\cal N}(\rho)) + o(n)$ qubits per
state, with the property that the original source output
can be recovered with high fidelity.
Schumacher's theorem \cite{JS,S} shows that the dimension needed for
asymptotically faithful encoding of a quantum source is equal to the
entropy of the density matrix of the source;
this gives the upper bound on $H({\cal N}^{\otimes n}(\pi_{T_n}))$
For the proof of the $\geq$ direction of Eq.~(\ref{nlimit}),
we need the theorem of Hausladen et al.~\cite{Haus} that the classical
capacity of signals transmitting pure quantum states is the entropy of
the density matrix of the average state transmitted (this is
a special case of Holevo's formula~(\ref{Holevo})).
We give a communication protocol which transmits
a classical message containing $n H({\cal N}(\rho)) - o(n)$ bits using
pure states.  By applying the theorem of Hausladen et al.\ to this
communication protocol, we deduce a lower bound on the entropy
${\cal N}^{\otimes n}(\pi_{T_n})$.

{\noindent {\bf Proof:}}
We first need some
notation.  Let the eigenvalues and eigenvectors of $\rho$ be
$\lambda_j$ and $\ket{v_j}$, with $1 \leq j \leq d$.
Let the noisy channel $\cal N$ map a $d$-dimensional space to
a $d_{\mathrm{out}}$-dimensional space.
Choose a Krauss representation for ${\cal N}$, so that
\[
{\cal N}(\sigma) = \sum_{k=1}^{c} A_k \sigma A_k^\dag \,,
\]
where $c \leq d^2$ and
$\sum_{k=1}^{c} A_k^\dag A_k = I$.
Then we have
\[
{\cal N}(\rho) = \sum_{j=1}^d \sum_{k=1}^{c} \lambda_j A_k \proj{v_j}
A_k^\dag.
\]
We let
\beq
\ket{u_{j,k}} = \frac{1}{\Big| A_k \ket{v_j}\Big|}  A_k \ket{v_j}
\eeq
and
\beq
\mu_{j,k} = \Big|A_k \ket{v_j}\Big|^2
\eeq
so that
\beq
{\cal N}(\rho) = \sum_{j=1}^d  \sum_{k=1}^{c}
\lambda_j \mu_{j,k} \proj{ u_{j,k} }\,.
\eeq

We need notation for the eigenstates and
eigenvalues of ${\cal N}(\rho)$.
Let these be $\ket{w_k}$ and $\omega_k$, $1 \leq k \leq d_{\mathrm{out}}$.
Finally, we define the probability $p_{jk}$, $1 \leq j \leq d$,
$1 \leq k \leq d_{\mathrm{out}}$, by
\beq
p_{jk} = \bra{w_k}\,{\cal N}\big(\ket{v_j}\bra{v_j} \big) \,\ket{w_k}.
\label{def-p}
\eeq
This is the probability that if the eigenstate $\ket{v_j}$ of $\rho$ is
sent through the channel ${\cal N}$ and measured in the eigenbasis
of ${\cal N}(\rho)$, that the eigenstate $\ket{w_k}$ will be observed.
Note that
\bea
\sum_j \lambda_j p_{jk} &=& \nonumber
\bra{w_k}\,{\cal N}\Big(\sum_j \lambda_j \ket{v_j}\bra{v_j} \Big)
\,\ket{w_k}\\
&=& \label{define-pjk}
\bra{w_k}\,{\cal N}(\rho) \,\ket{w_k}\\
&=& \omega_k \nonumber
\eea

We now define the typical subspace $T_{n,\delta,\rho}$.
Most previous papers on quantum information theory have dealt with
entropy typical subspaces.  We use frequency typical subspaces,
which are similar, but have properties that make the
proof of this lemma somewhat simpler.

A frequency typical subspace of ${\cal H}^{\otimes n}$ associated with
the density matrix $\rho \in {\cal H}$ is defined as the subspace
spanned by certain eigenstates of $\rho^{\otimes n}$.
We assume that $\rho$ has all positive eigenvalues.  (If it has some zero
eigenvalues,
we restrict to the support of $\rho$, and find the corresponding
typical subspace of $\supp(\rho)^{\otimes n}$, which will now
have all positive
eigenvalues.)
The eigenstates of $\rho^{\otimes n}$ are tensor product
sequences of eigenvectors of $\rho$, that is,
$\ket{v_{\alpha_1}}\otimes \ket{v_{\alpha_2}}\otimes \ldots\otimes
\ket{v_{\alpha_n}}$.
Let $\ket{s}$ be one of
these eigenstates of $\rho^{\otimes n}$.
We will say $\ket{s}$ is {\em frequency typical} if each
eigenvector $\ket{v_j}$
appears in the sequence $\ket{s}$ approximately
$n \lambda_j$ times.  Specifically,
an eigenstate $\ket{s}$ is {\em $\delta$-typical} if
\beq
\Big|\N_{\ket{v_j}}(\ket{s})-\lambda_j n\Big| < \delta n
\eeq
for all $j$; here $\N_{\ket{v_j}}(\ket{s})$ is the number of times that
$\ket{v_j}$ appears in $\ket{s}$.  The {\em frequency typical subspace}
$T_{n,\delta,\rho}$
is the subspace of ${\cal H}^{\otimes n}$ that is spanned by
all $\delta$-typical eigenvectors $\ket{s}$ of $\rho^{\otimes n}$.

We define $\Pi_T$ to be the projection onto the subspace $T$, and
$\pi_T$ to be this projection normalized to have trace $1$, that is,
$\pi_T = \frac{1}{\dim T}\Pi_T$.

 From the theory of typical sequences \cite{C-K}, for any density matrix
$\sigma$, any $\epsilon > 0$ and $\delta > 0$, one can choose $n$
large enough so that
\begin{enumerate}
\item
$\Tr \ \Pi_{T_{n,\delta,\sigma}} \, \sigma^{\otimes n}\,
\Pi_{T_{n,\delta,\sigma}} > 1-\epsilon$.
\item
The eigenvalues $\lambda$ of $\Pi_{T_{n,\delta,\sigma}}\,  \sigma^{\otimes n}\,  \Pi_{T_{n,\delta,\sigma}}$ satisfy
\[
2^{-n (H(\sigma)+\delta')} \leq \lambda \leq
2^{-n (H(\sigma)-\delta')}\ ,
\]
where $\delta' = \delta d \log (\lambda_{\max} / \lambda_{\min})$, and
$\lambda_{\max}$ ($\lambda_{\max}$) is the maximum (minimum) eigenvalue
of $\sigma$.
\item
$(1-\epsilon) 2^{n (H(\sigma) - \delta' )} \leq
\dim T_{n,\delta,\sigma} \leq 2^{n (H(\sigma) + \delta')}$.
\end{enumerate}
The property (1) follows from the law of large numbers, and (2), (3) are
straightforward consequences of (1) and the definition of typical subspace.

We first prove an upper bound that for all $\delta_1$, and for sufficiently
large $n$,
\beq
 \frac{1}{n} H({\cal N}^{\otimes n}(\pi_{T_{n,\delta_1,\rho}}))
<  H({\cal N}(\rho)) + C \delta_1\, .
\eeq
for some constant $C$.
We will do this by showing that for any $\epsilon$, there is an $n$
sufficiently large such that we can take a typical subspace
$T_{mn,{\delta_2},{\cal N}(\rho)}$ in
${\cal H}_{\mathrm{out}}^{\otimes n}$ and
project $m$ signals from a source with density matrix
${\cal N}^{\otimes n}(\pi_{T_{n,\delta_1,\rho}})$ onto it, such that the
projection has fidelity $1-\epsilon$ with the original output of
the source.
Here, $\delta_2$ (and $\delta_3$, $\delta_4$) will be a linear function
of $\delta_1$ (with the constant depending on $\sigma$, ${\cal N}$).
By projecting the source on $T_{mn,\delta_2, {\cal N}(\rho)}$, we are
performing Schumacher compression of the source.
 From the theorem on possible rates for
Schumacher compression (quantum source coding) \cite{JS,S}, this implies that
\beq
H({\cal N}^{\otimes n}(T_{n,\delta_1,\rho})) \leq
\lim_{m\rightarrow \infty} \frac{1}{m}
\log \dim T_{nm,{\delta_2},{\cal N}(\rho)}\, .
\eeq
The property (3) above for typical subspaces then implies the result.

Consider the following process.  Take a typical eigenstate
\[
\ket{s} = \ket{v_{\alpha_1}}\otimes \ket{v_{\alpha_2}}
\otimes \ldots \otimes \ket{v_{\alpha_n}}
\]
of $T_{n,\delta,\rho}$.
Now, apply a Krauss element $A_k$ to each symbol $\ket{v_{\alpha_j}}$
of $\ket{s}$, with element
$A_k$ applied with probability $\big| A_k \ket{v_{\alpha_j}} \big| ^2$.
This takes
\beq
\ket{s} = \bigotimes^n_{j=1} \ket{v_{\alpha_j}}
\eeq
to one of $c^{n}$ possible states $\ket{t}$.  Each
state is associated with a probability of reaching it; in particular,
the state
\beq
\label{define-t}
\ket{t} = \bigotimes^n_{j=1} \ket{u_{\alpha_j,\beta_j}}
\eeq
is produced with probability
\beq
\label{define-tau}
\tau = \prod_{j=1}^n \mu_{\alpha_j,\beta_j}.
\eeq

Notice that, for any $\ket{s}$, if
the $\ket{t_z}$ and $\tau_z$ are defined as in
Eqs. (\ref{define-t}) and (\ref{define-tau}), then
\beq
{\cal N}^{\otimes n}(\proj{s}) = \sum_{z=1}^{c^{n}} \tau_z \proj{t_z},
\eeq
where the sum is over all $\ket{t}$ in Eq.~(\ref{define-t}).

We will now see what happens when
$\ket{t_z}$ is projected onto a typical subspace
$T_{n, {\delta_2}, {\cal N}(\rho)}$ associated with
${\cal N}(\rho)^{\otimes n}$.  We get that the fidelity of this projection is
\beq
\bra{t_z}\Pi_{T_{n, {\delta_2}, {\cal N}(\rho)}}\ket{t_z} =
\sum_{\ket{r} \in T_{n,{\delta_2},{\cal N}(\rho)}}
\braket{r}{t_z} \braket{t_z}{r},
\eeq
where the sum is taken over all $\delta_2$-typical eigenstates $\ket{r}$ of
${\cal N}(\rho)^{\otimes n}$.
Now, we compute the average fidelity
(using the probability distribution $\tau$)
over all states $\ket{t_z}$ produced from a
given $\delta_1$-typical eigenstate $\ket{s} = \otimes_j \ket{v_{\alpha_j}}$:
\bea
\nonumber
\sum_z \tau_z\bra{t_z}\Pi_{T_{n, {\delta_2}, {\cal N}(\rho)}}\ket{t_z} &=&
\sum_{z=1}^{c^n}
\sum_{\ket{r} \in T_{n,{\delta_2},{\cal N}(\rho)}}
\tau_z \braket{r}{t_z} \braket{t_z}{r} \\
&=&
\nonumber
\sum_{\ket{r} \in T_{n,{\delta_2},{\cal N}(\rho)}}
\bra{r}\,{\cal N}^{\otimes n}(\proj{s}) \,\ket{r} \\
&=&\hspace*{-2em}
\sum_{\ket{r_z}= \bigotimes_{j=1}^n \ket{w_{\gamma_{z,j}}}
\atop
\ket{r_z} \in  T_{n,{\delta_2},{\cal N}(\rho)}\ \ \ \ }
\ \prod_{j=1}^n p_{\alpha_j,\gamma_{z,j}}.
\label{t-projection}
\eea
Here the last step is an application of Eq.~(\ref{def-p}).
The above quantity has a completely classical interpretation; it is
the probability that if we start with the $\delta_1$-typical sequence
$\ket{s}=\otimes\ket{v_{\alpha_j}}$, and take
$\ket{v_\alpha}$ to $\ket{w_\gamma}$ with
probability $p_{\alpha\gamma}$, that we end up with a ${\delta_2}$-typical
sequence of the $\ket{w_\gamma}$.

We will now show that the projection onto $T_{n, \delta_2, {\cal N}(\rho)}$
of the average state $\ket{t_z}$
generated from a $\delta_1$-typical eigenstate
$\ket{s}$ of $\rho^{\otimes n}$ has expected trace at least $1-\epsilon$.
This will be needed for the lower bound, and a similar result, using
the same calculations, will be used for the upper bound.
We know that the original sequence $\ket{s}$ is $\delta_1$-typical,
that is, each of the eigenvectors $\ket{v_j}$ appears approximately
$n\lambda_j$ times.  Now, the process of
first applying $A_k$ to each of the symbols,
and then projecting the result onto the eigenvectors
of ${\cal N}(\rho)^{\otimes mn}$,
takes $\ket{v_j}$ to $\ket{w_k}$ with probability $p_{jk}$.
We start with a $\delta_1$-typical sequence $\ket{s}$, so we have
\beq
\N_{\ket{v_j}}(\ket{s}) = (\lambda_j + \Delta_j) mn
\eeq
where $| \Delta_j | < \delta_1$.
Taking the state $\ket{s} = \bigotimes_j \ket{v_j}$ to
$\ket{r} = \bigotimes_k \ket{w_k}$, and using Eq.~(\ref{define-pjk}), we get
\bea
\nonumber
\E \left( \N_{\ket{w_k}}(\ket{r})\right)
 &=& (\omega_k + \sum_j \Delta_j p_{jk}) mn \\
&= & (\omega_k + \Delta_k' )mn
\eea
where $\Delta_k' \leq d \delta_1$.  The quantity
$\N_{\ket{w_k}}(\ket{r})$ is determined by the sum of $mn$
independent random variables whose values are either $0$ or $1$.
Let the expected average of these variables be
$\mu_k = \omega_k + \Delta_k'$.  Chernoff's bound \cite{AS}
says that for such a variable $X$ which is the sum of $N$ independent
trials, and $\mu N$ is the expected value of $X$,
\begin{eqnarray*}
\Pr[X - \mu N < -a ] &<& e^{-2a^2/N},
\\
\Pr[X - \mu N > a ] &<& e^{-2a^2/N}.
\end{eqnarray*}
Together, these bounds show that
\begin{equation}
\Pr[ \,|\N_{\ket{w_k}}(\ket{r}) - (\omega_k + \Delta_k')mn  |\,
< \delta mn] < 2 e^{-2\delta^2 mn}.
\end{equation}
If we take
$\delta_2 = (d+1)\delta_1$,  then by Chernoff's bound,
for every $\epsilon$ there are
sufficiently large $mn$ so that $\ket{r}$ is $\delta_2$-typical with
probability $1-\epsilon$.

Now, we are ready to complete the upper bound argument.
We will be using the theorem about
Schumacher compression \cite{JS,S} that if,
for all sufficiently large $m$,
we can compress $m$ states from a memoryless source emitting an ensemble
of pure states
with density matrix $\sigma$
onto a Hilbert space of dimension $m H$, and recover them with
fidelity $1-\epsilon$, then $H(\sigma) \leq H$.

We first need to specify a source with density matrix
${\cal N}^{\otimes n}(\pi_{T_{n, \delta_1, \rho}})$.
Taking a random $\delta_1$-typical
eigenstate $\ket{s}$ of $\rho^{\otimes n}$ (chosen uniformly from all
$\delta_1$-typical eigenstates), and premultiplying each of the tensor
factors $\ket{v_{\alpha_j}}$
by $A_k$ with the probability $\big| A_k \ket{v_{\alpha_j}} \big|$
to obtain a vector $\ket{t}$, gives us the desired source
with density matrix ${\cal N}^{\otimes n}(\pi_{T_{n,\delta_1, \rho}})$.
We next project a sequence of $m$ outputs from this
source onto the typical subspace $T_{mn,{\delta_2}, {\cal N}(\rho)}$.  Let us
analyze this process.  First, we will specify a sequence
$\ket{\bar{s}}$ of $m$ particular
$\delta_1$-typical eigenstates
$\ket{\bar{s}} = \ket{s_1}\ket{s_2}\cdots\ket{s_m}$.
Because each of the
components $\ket{s_i}$ of this state $\ket{\bar{s}}$ is $\delta_1$-typical,
$\ket{\bar{s}}$ is a
$\delta_1$-typical eigenstate of $\rho^{\otimes mn}$.  Consider the ensemble
of states $\ket{t}$ generated from any particular $\delta_1$-typical
$\ket{\bar{s}}$ by applying the $A_k$
matrices to $\ket{\bar{s}}$.  It suffices to show that this
ensemble can be projected onto
$T_{mn, {\delta_2}, {\cal N}(\rho)}$ with fidelity $1-\epsilon$;
that is, that
\beq
\sum_k \bra{\bar{s}} \otimes_k A_k^\dag \Pi_{T_{n,\delta_1,\rho}} \otimes_k A_k
\ket{\bar{s}} \geq 1-\epsilon.
\eeq
This will prove the theorem, as by averaging over all $\delta_1$-typical
states $\ket{\bar{s}}$ we obtain a source with density matrix
${\cal N}^{\otimes n}(\pi_{T_{n,\delta_1,\rho}})$ whose projection
has average fidelity $1-\epsilon$.  This implies, via the theorems on
Schumacher compression, that
\begin{eqnarray}
H ( {\cal N}^{\otimes n}(\pi_{T_{n,\delta_1, \rho}})) &\leq&
\lim_{m \rightarrow \infty}
\frac{1}{m} \dim \,  T_{mn,{\delta_2}, {\cal N}(\rho)} \nonumber \\
&\leq&
n (H({\cal N}(\rho)) + \delta_3)
\end{eqnarray}
where $\delta_3 = \delta_2 d_{\mathrm{out}} \log(\omega_{\max}/\omega_{\min})$;
here $\omega_{\max}$ ($\omega_{\min}$) is the maximum (minimum) non-zero
eigenvalue of ${\cal N}(\rho)$.
If we let $\delta_1$ go to 0 as $n$ goes to $\infty$, we obtain the
desired bound.  For this argument to work, we need to make sure that
$\epsilon$ is bounded independently of $\ket{\bar{s}}$; this follows
from the Chernoff bound.

We need now only show that the projection of the states $\ket{t}$ generated
from $\ket{s_1} \cdots \ket{s_m}$ onto the typical subspace
$T_{mn, {\delta_2}, {\cal N}(\rho)}$ has trace at least $1-\epsilon$.
We know that the original sequence $\ket{\bar{s}}$ is $\delta_1$-typical,
that is, each of the eigenvectors $\ket{v_i}$ appears approximately
$mn\lambda_i$ times.  Thus, the
same argument using the law of large numbers that
applied to Eq.~(\ref{t-projection}) also holds here,
and we have shown the upper bound for Lemma \ref{typical-lemma}.

We now give the proof of the lower bound.  We use the same notation
and some of the same ideas and machinery as in our proof of the upper bound.
Consider the distribution of $\ket{t_z}$ obtained
by first picking a random typical eigenstate $\ket{s}$ of $\rho^{\otimes n}$,
and applying a matrix $A_k$ to each symbol of $\ket{s}$, with $A_k$ applied
to $\ket{v_j}$ with probability $\big| A_k \ket{v_j}\big|^2$.
This gives an ensemble
of quantum states $\ket{t_z}$ with
associated probabilities $\tau_z$ such that
\beq
{\cal N}^{\otimes n}(\pi_{T_{n,\delta_1,\rho}}) =
\sum_{z=1}^{c^n} \tau_z \proj{t_z}.
\eeq
The idea for the lower bound is to choose randomly a set $T$ of
size $W = n ( H(\rho) - \delta_4)$ from the
vectors $\ket{t_z}$, according to the probability distribution
$\tau_z$.
We take $\delta_4 = C\delta_1$ for some constant $C$ to be determined
later.
We will show that with high probability (say, $1-\epsilon_2$)
the selected set $T$ of $\ket{t_z}$
vectors satisfy the
criteria of Hausladen et al \cite{Haus}
for having a decoding observable that correctly identifies a state $\ket{t_z}$
selected at random with probability $1-\epsilon$.  This means that these
states can be
used to send messages with rate $n (H(\rho)-\delta_4) (1-2\epsilon)$,
showing that
the density matrix of their equal mixture
$\pi_T = \frac{1}{|T|} \sum_{z \in T}\proj{t_z}$
has entropy at least
$n (H(\rho) -\delta_4))(1 -2 \epsilon)$.  However, the weighted average of
these density matrices $\pi_T$ over all sets $T$ is
${\cal N}^{\otimes n}(\pi_{T_{n,\delta_1,\rho}}) = \sum_z \tau_z \proj{t_z}$,
where each $\pi_T$ is weighted according to its probability of appearing.
By concavity of von Neumann entropy,
$H({\cal N}^{\otimes n}(\pi_{T_{n,\delta_1,\rho}})) \geq n (H(\rho) -\delta_4)
(1 -2 \epsilon)(1-\epsilon_2)$.  By amking $n$ sufficiently large,
we can make $\epsilon$, $\epsilon_2$, and $\delta_4$ arbitrarily small,
and so we are done.

The remaining step is to give the proof that with high probability
a randomly
chosen set of size $W$ of the $\ket{t_z}$ obeys the criterion of Hausladen
et al.  The Hausladen et al.\ protocol for decoding \cite{Haus}
is first to project onto a subspace, for which we will use the typical
subspace $T_{n, \delta_2, {\cal N}(\rho)}$, and
then use the square root measurement on the projected vectors.
Here, the square root measurement corresponding to vectors
$\ket{v_1}$, $\ket{v_2}$, $\cdots$  is the POVM with elements
\[
\phi^{-1/2} \proj{v_i} \phi^{-1/2}
\]
where
\[
\phi = \sum_{i} \proj{v_i}.
\]
Here, we use $\ket{v_i} = \Pi_{T_{n,\delta_w,{\cal N}(\rho)}} \ket{t_i}$.
Hausladen et al.\ \cite{Haus} give a criterion for the projection onto
a subspace followed by the square root measurement
to correctly identify a state chosen at random from the states
$\ket{t_z} \in T$.
Their theorem only gives the expected probability of error,
but the proof can easily be modified to show that the
probability of error $P_{E,i}$ in decoding the $i$'th vector, $\ket{t_i}$,
is at most
\beq
P_{E,i} \leq 2(1-S_{ii}) +  \sum_{j \neq i}S_{ij}S_{ji},
\label{Haus-criterion}
\eeq
where $S_{ii} = \bra{t_i} \Pi_{T_{n, \delta_2, {\cal N}(\rho)}}\ket{t_i}$ and
$S_{ij} = \bra{t_i} \Pi_{T_{n, \delta_2, {\cal N}(\rho)}}\ket{t_j}$.

We have already shown that the expectation of the first term of
(\ref{Haus-criterion}), $1-S_{ii}$, is small, for
$\ket{t_i}$ obtained from any typical eigenstate $\ket{s}$ of
$\rho^{\otimes n}$.
We need to give an estimate for the second term of (\ref{Haus-criterion}).
Taking expectations over all the $\ket{t_z}$, $z \neq i$, we obtain,
since all the $\ket{t_z}$ are chosen independently,
\beq
\E \Big( \sum_{j \neq i}S_{ij}S_{ji}\Big) = (W-1) \,\sum_{z=1}^{c^n}
\tau_z \Big|\bra{t_i}\Pi_{T_{n, \delta_2, {\cal N}(\rho)}}\ket{t_z}\Big|^2
\eeq
where $W$ is the number of random codewords $\ket{t_z}$ we choose randomly.
We now consider a different probability distribution on the $\ket{t_z}$,
which we call $\tau_z'$.
This distribution is obtained by first choosing
an eigenstate $\ket{s}$ of $\rho^{\otimes n}$ with probability proportional
to its eigenvalue (rather than choosing uniformly among $\delta$-typical
eigenstates of $\rho^{\otimes n})$, and
then applying a Krauss element $A_k$ to each of its symbols to obtain
a word $\ket{t}$ (as before, $A_k$ is applied to $\ket{v_j}$ with probability
$\big|A_k\ket{v_j}\big|^2$).
Observe that $\tau_z < 2^{2 \delta' n}\tau_z'$, where
$\delta' = d \delta \log(\lambda_{\max}/\lambda_{\min})$.  This
holds because the difference between the two distributions $\tau$ and $\tau'$
stems from the probability with which an eigenstate $\ket{s}$ of
$\rho^{\otimes n}$ is chosen; from the properties of typical subspaces,
the eigenvalue of every typical eigenstate $\ket{s}$ of
$\rho^{\otimes n}$ is no more than $2^{- n ( H(\rho) - \delta')}$, and
the number of such eigenstates is at most
$2^{n (H(\rho)+\delta')}$.  Thus, we have
\bea
\nonumber
\E \left( \sum_{j \neq i}S_{ij}S_{ji} \right) &\leq&
W \sum_z \tau_z
\bra{t_i}\Pi_{T_{n, \delta_2, {\cal N}(\rho)}}\ket{t_z}
\bra{t_z}\Pi_{T_{n, \delta_2, {\cal N}(\rho)}}\ket{t_i}\\
\nonumber
&\leq & W
2^{2 \delta' n}
\sum_z \tau_z'
\bra{t_i}\Pi_{T_{n, \delta_2, {\cal N}(\rho)}}\ket{t_z}
\bra{t_z}\Pi_{T_{n, \delta_2, {\cal N}(\rho)}}\ket{t_i}\\
\nonumber
&=& W
2^{2 \delta' n}
\bra{t_i}\Pi_{T_{n, \delta_2, {\cal N}(\rho)}}
{\cal N}(\rho)^{\otimes n}
\Pi_{T_{n, \delta_2, {\cal N}(\rho)}}\ket{t_i}\\
&\leq & W
2^{2 \delta' n}
2^{- n (H(\rho)-\delta_3)}
\eea
where the last inequality follows from property (2) of typical subspaces,
which gives a bound on the maximum
eigenvalue of
$\Pi_{T_{n, \delta_2, {\cal N}(\rho)}} {\cal N}(\rho)^{\otimes n}
\Pi_{T_{n, \delta_2, {\cal N}(\rho)}}$.
Thus, if we make $W = 2^{n (H(\rho) - 2\delta' - \delta_3 - \delta)}$,
we have the desired inequality (\ref{Haus-criterion}),
and the proof of Lemma \ref{typical-lemma} is
complete.

We used frequency-typical subspaces rather than entropy-typical
subspaces in the proof of Lemma \ref{typical-lemma}; this
appears to be the most natural method of proof.
Holevo \cite{newHolevo} has found a more direct proof of
Lemma \ref{typical-lemma}, which also uses frequency-typical subspaces.
Frequency-typical
sequences are commonly used in classical information theory, although
they have not yet seen much use in quantum information theory,
possibly because the quantum information community has not had much
exposure to them.
One can ask whether Lemma \ref{typical-lemma} still holds for entropy-typical
subspaces.  This is not only a natural question, but might also
be a method of extending Lemma \ref{typical-lemma} to the case where
$\supp(\rho)$ is a countable-dimension Hilbert space, a case where the
method of frequency-typical subspaces does not apply.
The difficulty with using entropy-typical subspaces in our current proof
is that an eigenstate $\ket{s}$ of $\rho^{\otimes n}$
which is entropy-typical but not frequency-typical will in general not
be mapped to a mixed state ${\cal N}(\proj{s})$ having most of its mass
close to the typical eigenspace of ${\cal N}(\rho)^{\otimes n}$.
This means that the Schumacher compression argument is no longer valid.
One way to fix the problem
is to require an extra condition on the eigenvectors of
the typical subspace which implies that most of their mass is indeed
mapped somewhere close to the typical eigenspace of
${\cal N}(\rho)^{\otimes n}$.
We have found such a condition (automatically satisfied by frequency-typical
eigenvectors), and believe this may indeed be useful for
studying the countable-dimensional case.

\subsection{Proof of the Upper Bound}
We prove an upper bound of
\begin{equation}
C_E \leq \max_{\rho \in {\cal H}_{\rm{in}}} \ \
 H(\rho) + H({\cal N}(\rho)) - H({\cal N} \otimes {\cal I}(\Phi_\rho))
\label{upper-bound}
\end{equation}
where $\Phi_\rho$ is a purification of $\rho$.

As in the proof of the lower bound, this proof works by first proving
the result in a special case and then using this special case to obtain
the general result.  Here, the special case is when Alice's protocol
is restricted to encode the signal using a unitary transformation of
her half of the entangled state $\phi$.  This special case is proved
by analyzing the possible protocols, applying the capacity formula
(\ref{Holevo}) of Holevo and Schumacher and Westmoreland~\cite{Holevo,SW},
and then applying several entropy inequalities.

First, consider a channel ${\cal N}$ with entanglement-assisted capacity $C_E$.
By the definition of entanglement-assisted capacity, for every
$\epsilon$, there is a protocol that uses the channel ${\cal N}$ and some block
length $n$, that achieves capacity
$C_E - \epsilon$, and that does the following:

Alice and Bob start by sharing a pure entangled state $\phi$,
independent of the classical data Alice wishes to send. (Protocols where they
start with a mixed
entangled state can easily be simulated by ones starting with a
pure state, although possibly at the cost of additional entanglement.)
Alice then performs some superoperator ${\cal A}_x$ on her half of $\phi$
to get $({\cal A}_x \otimes {\cal I})(\phi)$, where ${\cal A}_x$ depends on
the classical
data $x$ she
wants to send.  She then sends her half of ${\cal A}_x(\phi)$ through the
channel ${\cal N}^{\otimes n}$ formed by the tensor product of $n$ uses of the
channel ${\cal N}$.
Bob then possibly waits until he receives many of these states
$({\cal N}^{\otimes n}\otimes {\cal I})({\cal A}_x\otimes {\cal I})(\phi)$, and applies
some decoding procedure to them.

This follows from
the definition of entanglement-assisted capacity (\ref{framework})
using only forward communication.  Without feedback from Bob to Alice,
Alice can do no better
than encode all her classical information at once, by applying a single
classically-chosen completely positive map ${\cal A}_x$ to her half of the
entangled state $\phi$, and then send it to Bob through the noisy
channel ${\cal N}^{\otimes n}$. (If, on the contrary, feedback
were allowed, it might be advantageous to use a protocol
requiring several rounds of communication.)  Note that the
present formalism includes situations where Alice doesn't use
the entangled state $\phi$ at
all, because the map ${\cal A}_x$ can completely discard
all the information in $\phi$.

In this section, we assume that ${\cal A}_x$ is a unitary
transformation ${\cal U}_x$. Once we have derived an upper bound
assuming that Alice's transformations are unitary, we will use this upper
bound to show that allowing her to use non-unitary transformations does
not help her. This is proved using the strong subadditivity property
of von Neumann entropy; the proof (Lemma \ref{show-unitary})
will be deferred to the next section.

The next step
in our proof is to apply the Holevo formula, Eq.~(\ref{Holevo}), to the
tensor product channel ${\cal N}^{\otimes n}$.  Let
$\hat{\cal N} = {\cal N}^{\otimes n}$
denote the tensor product of many uses of the channel. For the $x$th
signal state, Alice sends her half of $({\cal U}_x \otimes {\cal I})(\phi)$
through the channel $\hat{\cal N}$, and Bob receives
$(\hat{\cal N}\otimes {\cal I})({\cal U}_x \otimes {\cal I})(\phi)$.
Bob's state can be divided into two parts. The
first of these
is his half of $\phi$, which, after Alice's part is traced out, is
always in state $\Tr_A (\phi)$. The second part is the state Alice sent
through the channel, which, after Bob's part is traced out, is in state
$\hat{\cal N}(\rho_x)$ where $\rho_x = \Tr_B ({\cal U}_x(\phi))$.
Bob is trying to decode
information from the output of many blocks, each containing $n$ uses of
the channel, together with his half of the associated entangled states,
i.e., from many blocks of the form
$(\hat{\cal N}\otimes {\cal I})({\cal U}_x \otimes {\cal I})(\phi)$.
Since these blocks are not entangled each other,
the Holevo-Schumacher-Westmoreland theorem \cite{Holevo,SW} applies,
and the capacity is given by formula (\ref{Holevo}),
considering these blocks to be the signal states.
The first term of formula
(\ref{Holevo}) is the entropy of the average block, and this is
bounded by
\beq
H(\hat{\cal N} (\sum_x p_x\rho_x) ) + H(\rho_x).
\label{termone}
\eeq
The first
term in (\ref{termone}) is the entropy of the average state that Bob
receives through the channel, i.e., $\hat{\cal N} ({\cal U}_x (\Tr_B \phi))$,
and the second term is the entropy of the state that Bob retained all
the time, i.e., $\Tr_A \phi$.
That the sum of the two terms is a bound for the entropy follows from
the subadditivity property of
von Neumann entropy that the entropy of a joint system is
bounded from above by the sum of the entropies of the two systems \cite{NC}.
We can use $H(\rho_x)$ for the second term because Alice is using a unitary
transformation to produce $\rho_x$ from her half of the entangled state $\phi$
she shares with Bob, so the entropy $H(\rho_x) = H(\Tr_A \phi)$ is the same
for all $x$.
Since we assume that Alice and Bob share a pure quantum state, the
entropy of Bob's half is the same as the entropy of Alice's half.
Although this is not the most obvious expression for this second
term of (\ref{termone}),
it will facilitate later manipulations.

The second term of formula (\ref{Holevo}) is the average entropy of the
state Bob receives, and this is
\beq
\sum_x p_x H((\hat{\cal N}\otimes {\cal I})(\Phi_{\rho_x}))
\eeq
where $\Phi_{\rho_x}$ is a purification of $\rho_x$.  This formula holds
because Alice's and Bob's joint state after Alice's unitary transformation
${\cal A}_x$ is still a pure state, and so their joint state is a
purification of $\rho_x$.

We thus get
\beq
n(C_E-\epsilon) \leq H\left(\hat{\cal N} (\sum_x p_x\rho_x) \right) +
\sum_x p_x H(\rho_x)
- \sum_x p_x H(\hat{\cal N}\otimes {\cal I}(\Phi_{\rho_x})).
\eeq
However, by Lemma \ref{concavity-lemma}, that we
prove in the next section,
the last two terms in this formula are a concave function of $\rho_x$,
so we can move the sum inside these terms, and we get
\begin{equation}
C_E - \epsilon \leq \frac{1}{n}\left( H(\hat{\cal N} (\rho) ) + H(\rho)
-  H(\hat{\cal N}\otimes {\cal I}(\Phi_{\rho}))\right)
\label{CE-n-large}
\end{equation}
where
\[
\rho = \sum_x p_x \rho_x.
\]
Finally, the expression (\ref{CE}) for $C_E$ is additive
(this will be discussed in the next section), so that
\beq
C_E({\cal N}_1\otimes {\cal N}_2)=
C_E({\cal N}_1) +
C_E({\cal N}_2).
\eeq
Using this, we can set $n=1$ in Eq.~(\ref{CE-n-large}), thus
replacing $\hat{\cal N}={\cal N}^{\otimes n}$ by ${\cal N}$.
Since this equation holds for any $\epsilon>0$,
we obtain the desired formula (\ref{upper-bound}).

\subsection{Proofs of the Lemmas}

This section discusses three lemmas needed for the previous section.
The first of these shows that without loss of capacity, Alice can use a
unitary transform for encoding.  The next shows that the last two
terms of the formula for $C_E$ in Eq.~(\ref{CE}) are a convex function of
$\rho$.  The last lemma shows that the formula for $C_E$ is additive.
The first two lemmas use the property of strong subadditivity for
von Neumann entropy.
Originally, we also had a fairly complicated proof for the third lemma.
However, Prof.\ Holevo has pointed out that a much simpler proof (also
using strong subadditivity) was already in the literature,
and so we will merely cite it.

For the proofs of the first two lemmas in this section, we
need the strong subadditivity
property of von Neumann entropy \cite{LR,NC}.  This property says that if
$A$, $B$, and $C$ are quantum systems, then
\beq
H(\rho_{AB}) + H(\rho_{AC}) \geq H(\rho_{ABC}) + H(\rho_A).
\eeq
It turns out to be a surprisingly strong property.

We need to show
that if Alice uses non-unitary transformations~${\cal A}_x$,
then she can never do better than
the upper bound Eq.~(\ref{upper-bound}) we
derived by assuming that she uses only unitary transformations ${\cal U}_x$.
Recall that any non-unitary transformation ${\cal A}_x$ on a Hilbert
space ${\cal H}_{\rm{in}}$ can be performed by using a unitary transformation
${\cal U}_x$ acting on the Hilbert space ${\cal H}_{\rm{in}}$ augmented by
an ancilla space ${\cal H}_{\rm{anc}}$, and then tracing out the ancilla
space \cite{NC}.
We can assume that $\dim {\cal H}_{\rm{anc}} \leq (\dim {\cal H}_{\rm{in}})^2$.

What we will do is take the channel ${\cal N}$ we were given, that acts on
a Hilbert space ${\cal H}_{\rm{in}}$ and simulate it by a channel
${\cal N}'$
that acts
on a Hilbert space ${\cal H}_{\rm{in}} \otimes {\cal H}_{\rm{anc}}$ where
${\cal N}'$ first traces out ${\cal H}_{\rm{anc}}$ and then applies
${\cal N}$ to the residual state on ${\cal H}_{\rm{in}}$.  We can then perform
any transformation $S_x$ by performing a unitary operation ${\cal U}_x$
on ${\cal H}_{\rm{in}} \otimes {\cal H}_{\rm{anc}}$ and tracing
out ${\cal H}_{\rm{anc}}$.
Since we proved the formula Eq.~(\ref{upper-bound}) for unitary
transformations in the previous section, we can calculate $C_E$
by applying this formula
to the channel ${\cal N}'$.  What we show below is that the same
formula applied
to ${\cal N}$ gives a quantity at least as large.
\vspace{-.5ex}
\begin{lem}
\label{show-unitary}
Suppose that ${{\cal N}}$ and ${{\cal N}}'$ are related as
described above.  Let us define
\begin{equation}
C = \max_{\rho \in {\cal H_{\rm{in}}}} H(\rho) + H({\cal N}(\rho)) - H({\cal N} \otimes {\cal I}(\Phi_\rho))
\label{C}
\end{equation}
and
\begin{equation}
C' = \max_{\rho' \in {\cal H}_{\rm{in}} \otimes
{\cal H}_{\rm{anc}}} H(\rho') + H({\cal N}'(\rho'))
- H({\cal N}' \otimes {\cal I}(\Phi_{\rho'})).
\label{Cprime}
\end{equation}
Then $C \geq C'$.
\end{lem}

{\noindent {\bf Proof:}}
To avoid double subscripts in the following calculations, we now
rename our Hilbert spaces as follows.
Let
$A = {\cal H}_{\rm{in}}$;
$A' = {\cal H}_{\rm{anc}}$;
$B = {\cal H}_{\rm{out}}$;
and
$E = {\cal H}_{\rm{env}}$.
Let $\rho'$ maximize $C'$ in the above formula.  We let
$\rho = \Tr_{A'} \rho'$.
Since the channel ${\cal N}'$ was defined by first tracing out $A'$ and then
sending the resulting state through the channel ${\cal N}$, $\rho$ is the
density matrix of the state input to the channel ${\cal N}$ in the
protocol.

Clearly, the middle terms in the above two
formulae (\ref{C}) and (\ref{Cprime})
are equal, since ${\cal N}(\rho) = {\cal N}'(\rho')$.
We need to show that inequality holds for the first and last terms
in $C$ and $C'$; that is, we need to show
\beq
H(\rho) - H(({\cal N}\otimes {\cal I})(\Phi_\rho)) \geq
H(\rho') - H(({\cal N'}\otimes {\cal I})(\Phi_{\rho'})).
\eeq

\begin{figure}[tbp]
\epsfxsize=2.2cm
\hspace*{1.25in}\epsfbox{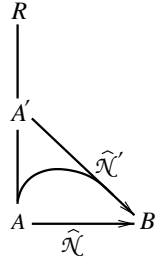}\medskip
\caption{In Lemma \ref{show-unitary},
$A$ is the input space for the original map ${\cal N}$.
$A \cup A'$ is the input space for the map ${\cal N}'$.
The output space for
both maps is $B$.  The space $R$ is a reference system used to purify
states in $A$ and $A'$.
}
\label{bigcef2}
\end{figure}

Recall, we have a noisy channel ${\cal N}$ that acts on Hilbert space $A$,
and a channel ${\cal N}'$ that acts on Hilbert space $A \otimes A'$
by tracing out $A'$ and then sending the resulting
state through ${\cal N}$.
We need to give
purifications $\Phi_\rho$ and $\Phi_{\rho'}$ of $\rho$ and
$\rho'$, respectively.
Note that we can take $\Phi_\rho = \Phi_{\rho'}$, since any
purification of $\rho'$ is also a purification of $\rho$ 
(see footnote 2).  Let us
take these purifications over a reference system ${\cal H_{\mathrm{ref}}}$
that we call $R$.
Consider the diagram in Figure \ref{bigcef2}.
In this figure, $\rho_A = \rho$, $\rho_{AA'} = \rho'$ and $\rho_{AA'R} =
\ket{\Phi_\rho}\bra{\Phi_\rho} =  \ket{\Phi_{\rho'}}\bra{\Phi_{\rho'}}$.
Then ${\cal N}$ maps the space $A$ to the space $B$ and
${\cal N}'$ maps the space $AA'$ to the space $B$ by tracing out $A'$ and
performing ${\cal N}$.

We have $H(\rho) = H(\rho_A) = H(\rho_{A'R})$, and
$H(\rho') = H(\rho_{AA'}) = H(\rho_R)$.  We also have
$H(({\cal N} \otimes {\cal I})(\Phi_\rho)) = H(\rho_{A'RB})$ and
$H(({\cal N}' \otimes {\cal I})(\Phi_{\rho'})) = H(\rho_{RB})$.

Thus,
\begin{eqnarray}
\nonumber
C - C'
&=& H(\rho) - H(({\cal N}\otimes {\cal I})(\Phi_\rho))
- H(\rho') + H(({\cal N}' \otimes {\cal I})(\Phi_{\rho'})) \\
&=& H(\rho_{A'R}) -  H(\rho_{A'RB}) -  H(\rho_R) +  H(\rho_{RB})\\
\nonumber
&\geq& 0
\end{eqnarray}
by strong subadditivity, and we have the desired inequality.

For the next lemma, we need to prove that
the function
\[
H(\rho) - H((\hat{\cal N} \otimes {\cal I})(\Phi_\rho))
\]
is concave in $\rho$.
\begin{lem}
\label{concavity-lemma}
Let $\rho_0$ and $\rho_1$ be two density matrices, and
let $\rho = p_0 \rho_0 + p_1 \rho_1$ be their weighted average. Then
\begin{eqnarray}
\label{concavity}
\nonumber
H(\rho) - H((\hat{\cal N} \otimes {\cal I})(\Phi_\rho)) &\geq&\!\! \phantom{+\ }
p_0(H(\rho_0) - H((\hat{\cal N} \otimes {\cal I})(\Phi_{\rho_0}))) \\
&& \!\! + \  p_1(H(\rho_1)
- H((\hat{\cal N} \otimes {\cal I})(\Phi_{\rho_1}))).
\end{eqnarray}
\end{lem}

\begin{figure}[tbp]
\epsfxsize=2.2cm
\hspace*{1.25in}\epsfbox{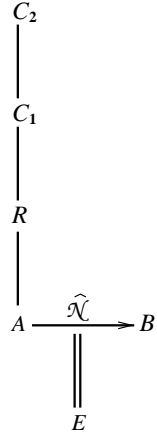}\medskip
\caption{ For Lemma \ref{concavity-lemma}, $A$ is a Hilbert space we
send through the channel
${\hat{\cal N}}$, and $B$ is the output space.  This mapping $\hat{\cal N}$
can be made unitary by adding an environment space $E$.   We let $R$ be
a reference system which purifies the systems $\rho_0$ and $\rho_1$ in $A$,
and $C_1$ and $C_2$ be two qubits purifying $AR$ as described in the text.
}
\label{bigcef3}
\end{figure}

\noindent{\bf Proof:}
We again give a diagram; see Figure \ref{bigcef3}.
Here we let the states be as follows:
$\rho_A = \rho = p_0 \rho_0 + p_1 \rho_1$, so $A$ is in the state $\rho$.
We let $R$ be a reference system with which we purify the states $\rho_0$
and $\rho_1$. Consider purifications
$\Phi_{0} = \proj{\phi_0}$ and $\Phi_{1} = \proj{\phi_1}$ of
$\rho_0$, $\rho_1$, respectively.
Then we have
\beq
\rho_{AR} = p_0 \ket{\phi_0}\bra{\phi_0} + p_1 \ket{\phi_1}\bra{\phi_1}.
\eeq
We now let $C_1$ and $C_2$ be qubits which tell whether the
system $A$ is in state $\rho_0$ or $\rho_1$, and we will purify the
system $\rho_{AR}$ in the system $ARC_1C_2$ in the following way:
\beq
\phi_{ARC_1C_2} =
\sqrt{p_0} \ket{\phi_0} \ket{0} \ket{0} +
\sqrt{p_1} \ket{\phi_1} \ket{1} \ket{1}.
\eeq
Tracing out $C_2$, we get that the state of $ARC_1$ is
\beq
\rho_{ARC_1} = p_0 \ket{\phi_{0}} \bra{\phi_{0}}
\otimes \ket{0}\bra{0}
+  p_1 \ket{\phi_{1}} \bra{\phi_{1}}
\otimes   \ket{1}\bra{1},
\eeq
so now $C_1$ can be thought of as a classical bit telling which of
$\Phi_0$ or $\Phi_1$ is the state of the system $AR$.
Note that we have the same expression after tracing out $C_2$.

Now, it's time for our
analysis.  We want to show equation (\ref{concavity}) above.
Notice that
\[
H(\rho) = H(\rho_A) = H(\rho_{RC_1C_2}),
\]
since $\rho_{ARC_1C_2}$ is in a pure state,
and
\[
H((\hat{\cal N} \otimes {\cal I})(\Phi_\rho)) = H(\rho_{BRC_1C_2}).
\]
Now, suppose we have a classical bit $C$ which tells whether a quantum
system $X$ is in state $\rho_0$ or $\rho_1$, with probability
$p_0$ and $p_1$ respectively.  The following formula
gives the expectation of the entropy of $X$ \cite{NC,Wehrl}
(this is analogous to the chain rule for the entropy of classical
systems):
\begin{eqnarray}
E(\rho_X) &=& p_0 H(\rho_0) + p_1 H(\rho_1) \nonumber \\
&=& H(\rho_{XC}) - H(\rho_C).
\label{classicalbit}
\end{eqnarray}
Using this formula (\ref{classicalbit}), we see that
\beq
\sum_{j=0}^1
p_j H((\hat{\cal N} \otimes {\cal I})(\Phi_{\rho_j}))
= H(\rho_{BRC_1}) - H(\rho_{C_1})
\eeq
and
\begin{eqnarray}
\sum_{j=0}^1 p_j H(\rho_j) &=& H(\rho_{AC_2}) - H(\rho_{C_2}) \nonumber
\\
&=& H(\rho_{RC_1}) - H(\rho_{C_2}).
\end{eqnarray}

Putting everything together, we get
\begin{eqnarray}
&& \hspace*{-.75in} H(\rho) - H((\hat{\cal N} \otimes {\cal I})(\Phi_\rho)) -
\sum_{j=0}^1 p_j
\left( H(\rho_j) - H((\hat{\cal N} \otimes {\cal I})(\Phi_{\rho_j})) \right)
\nonumber\\
&=&
H(\rho_{RC_1C_2}) - H(\rho_{BRC_1C_2})
- H(\rho_{RC_1}) + H(\rho_{BRC_1})
\label{endlem3}
\end{eqnarray}
which is positive by strong subadditivity.  To obtain (\ref{endlem3}), we
used the equality
$H(\rho_{C_1}) = H(\rho_{C_2})$, which holds by symmetry.  This concludes
the proof of Lemma \ref{concavity-lemma}.

The final lemma we need shows that we can set $n=1$ and replace
$\hat{\cal N}={\cal N}^{\otimes n}$ by ${\cal N}$
in Eq.~(\ref{CE-n-large}).
This follows from the fact that $C_E$ is additive, that is, if $C_E$ is
taken to be defined by Eq.~(\ref{CE}), then
\beq
C_E({\cal N}_1 \otimes {\cal N}_2) = C_E({\cal N}_1) + C_E({\cal N}_2).
\eeq
The $\geq$ direction is easy.  We originally had a rather unwieldy
proof for the
$\leq$ direction based on explicitly expanding the
formula for $C_E$ and differentiating; However,
A. Holevo has pointed out to us
that a much simpler proof is given in \cite{CandA}, so we will spare the
readers our proof.

\section{Examples of $C_E$ for Specific Channels}

In this section, we discuss the capacity of two specific channels:
the first is
the bosonic channel with attenuation/amplification and Gaussian noise,
given a bound on the average signal energy, and the second is the
qubit amplitude damping channel. Strictly speaking, we have not
yet shown that the formula (\ref{CE}) holds for the Gaussian bosonic
channel, as
we have not proved that it holds either given
an average energy constraint or
for continuous channels.  For channels with a linear constraint on the
average density matrix $\rho$, our proof applies unchanged, and yields
the result that the density matrix $\rho$ of (\ref{CE})
must be optimized over all density matrices satisfying this linear constraint.
We make no claims as to having proven the formula (\ref{CE}) for
continuous channels.
In fact, we suspect that there may be continuous quantum channels which
have a finite entanglement-assisted capacity, but where each of the terms
of the formula (\ref{CE}) is infinite for the optimal density matrix for
signaling.  The theory of entanglement-assisted capacity for continuous
channels is thus currently incomplete.

For the Gaussian channel with an average
energy constraint, all three terms of (\ref{CE}) must be finite,
since any bosonic state with finite energy has a finite entropy.
For this channel, (\ref{CE}) can be proven by approximating
the channel with a sequence of finite-dimensional
channels whose capacity we can show converges to the
capacity of the Gaussian channel.  We do this approximation by firstly
restricting the input to the channel a finite subspace, and secondly
projecting the output of the channel onto a finite subspace.  (In these
cases, the finite subspace can be taken to be that generated by the
first $k+1$ number basis states $\ket{n=0}$, $\ket{n=1}$, $\ldots$,
$\ket{n=k}$ defined later in this section.)

\subsection{Gaussian Channels}

The Gaussian channel is one of the most important continuous alphabet
classical channels, and we briefly review it here.  We describe the
classical complex Gaussian channel, as this is
most analogous to the quantum Gaussian channel.
For a detailed discussion of this channel see an information theory
text such as \cite{CT,C-K}.

A classical complex Gaussian channel $\rm{N}$ of noise $N$ is defined by
the mapping in the complex plane
\begin{equation}
{\rm N}: z \mapsto z' \, , \ \ \ z'\sim G_N(z'-z) \enspace,
\end{equation}
where the noise $G_N$ is a Gaussian of mean 0 and variance $N$, i.e.,
\begin{equation}
G_N(z) = \frac{1}{{\pi N}} e^{- |z|^2/N}.
\end{equation}
Without any further conditions, the capacity of this channel would be
unlimited, because we could choose an infinite subset of inputs
arbitrarily far apart so that the corresponding outputs are distinguishable
with arbitrarily small probability of error.
We add an additional constraint on average input signal power or energy,
say $S$.  That is, we require that the input distribution $W(x)$ satisfy
\beq
\int |z|^2 W(z) \, d^2 z \leq S.
\eeq
This complex Gaussian channel is equivalent to two parallel
real Gaussian channels.  It follows that the capacity of
the complex Gaussian channel with average input energy $S$ and noise $N$ is
\beq
C_{\mathrm{Shan}} = \log \left(1 + \frac{S}{N}\right),
\eeq
which is twice the capacity of a real Gaussian channel with
average input energy $S$ and noise $N$.

Before we proceed to discuss the quantum Gaussian channel, let us first
review some basic results from quantum optics.  In the quantum theory of
light, each mode of the electromagnetic field is treated as a quantum
harmonic oscillator whose commutation relations are the same as those
of $SU(1,1)$.  A detailed treatment of these concepts is available in
the book \cite{wal-mil}.
The Hilbert space corresponding to a mode is countably infinite.
A countable orthonormal basis for this space is the number basis of states
$\ket{n=j}$, $j = 0, 1, 2, \ldots $,
where the state $\ket{n=j}$ corresponds to
$j$ photons being present in the mode.

Another useful basis is that
of the coherent states of light.  Coherent states are defined for complex
numbers $\alpha$ as
\begin{eqnarray}
\ket{\alpha} &=& D(\alpha)\ket{0}\\
&=& e^{-|\alpha|^2/2} \sum_{j=0}^{\infty} \frac{\alpha^j}{\sqrt{j!}}\ket{n=j}
\end{eqnarray}
where $D(\alpha)$ is the unitary displacement operator
and $\ket{0} = \ket{n=0}$ is the vacuum state containing no photons.
The complex number
$\alpha$ corresponds to the complex field vector of a mode in the
classical theory of light.
If $\alpha = x + i p$, then $x$ is generally
called the position coordinate and
$p$ the momentum coordinate.
The displacement operator corresponds to displacing the complex number
labeling the coherent state, and multiplying by an associated phase, i.e.,
\beq
D(\alpha) \ket{\beta} = \ket{\alpha + \beta} e^{i\,\Im (\alpha \beta^*)}
\eeq
where $\Im$ takes the imaginary part of a complex number, i.e.,
$\Im(x+i y) = y$.

We also need thermal states, which are the equilibrium distribution
of the harmonic oscillator for a fixed temperature.  The thermal
state with average energy $S$ is the state
\begin{eqnarray}
\nonumber
T_S &= &\frac{1}{S+1}\sum_{j=0}^\infty
\left( \frac{S}{S+1}\right)^j \proj{n=j}\\
&=& \frac{1}{\pi S}\int e^{-|z|^2/S}\proj{z}\, d^2 z.
\end{eqnarray}
The entropy of the thermal state $T_S$ is
\beq
g(S) = (S+1)\log(S+1) - S \log(S)\, .
\eeq

We are now ready to define the quantum analog of the classical
Gaussian channel.
(See \cite{HW99} for a much
more detailed treatment of quantum Gaussian channels.)
Coherent states are an overcomplete basis, and
a quantum channel may be defined by its action
on coherent states.
We restrict our discussion to quantum Gaussian channels with one mode
and no squeezing, which are those most analogous to classical Gaussian
channels.
These channels have an attenuation/amplification
parameter $k$, and a noise parameter $N$.
The channel amplifies the
signal (necessarily introducing noise) if $k > 1$,
and attenuates the signal if $k < 1$.
Amplification/attenuation of the quantum state intuitively corresponds
to multiplying the average position and momentum coordinates by the
number $k^2$.  If this were possible for $k > 1$ without introducing any
extra noise, it would enable one to violate the Heisenberg uncertainty
principle and measure the position and momentum coordinates simultaneously
to any degree of accuracy by first amplifying the signal and then 
simultaneously measuring these coordinates with optimal quantum uncertainty.  
To ensure that the channel is a completely positive map, amplification 
thus must necessarily entail introduce extra quantum noise.  
The channel ${\cal N}$ with noise $N$ and 
attenuation/amplification
parameter $k$ acts on coherent states as
\begin{eqnarray}
\nonumber
{\cal N} (\proj{\alpha}) &=& D_{k^2 \alpha} T_N
D^\dag_{k^2 \alpha\phantom{{}+k^2-1}}
{\rm{\ \ \ \ for \ }} k \leq 1\\
{\cal N} (\proj{\alpha}) &=& D_{k^2 \alpha} T_{N+k^2-1}
D^\dag_{k^2 \alpha }
{\rm{\ \ \ \ for \ }} k \geq 1.
\end{eqnarray}

The entanglement-assisted capacity of Gaussian channels was
calculated in \cite{HW99}.  The density matrix $\rho$ maximizing $C_E$
is a thermal state of average energy $S$, and the entanglement-assisted
capacity is given by
\beq
C_E = g(S) + g(S') - g(\frac{D+S' -S -1}{2}) - g(\frac{D-S' +S -1}{2}).
\label{Gaussian-CE}
\eeq
Here $S$ is the average input energy; $S'$ is the average output energy:
\begin{eqnarray}
\nonumber
S' &=& k^2 S + N \phantom{ + k^2 - 1 }\, \  \mathrm{\,\ \ \ \ for \ } k \leq 1\\
S' &=& k^2 S + N + k^2 - 1\,  \ \mathrm{\ \ \ for \ } k \geq 1;
\end{eqnarray}
and
\begin{equation}
D = \sqrt{(S + S' + 1)^2 - 4 k^2 S(S+1)}.
\label{general-D}
\end{equation}
The first term of (\ref{Gaussian-CE}), $g(S)$, is the entropy of the
input; the second term, $g(S')$, is the entropy of the output; and the
remaining two terms of (\ref{Gaussian-CE}) are the entropy of a
purification of the thermal state $T_S$ after half of it has passed
through the channel.

\begin{figure}[tbp]
\epsfxsize=5in\hspace*{.5in}
\epsfbox{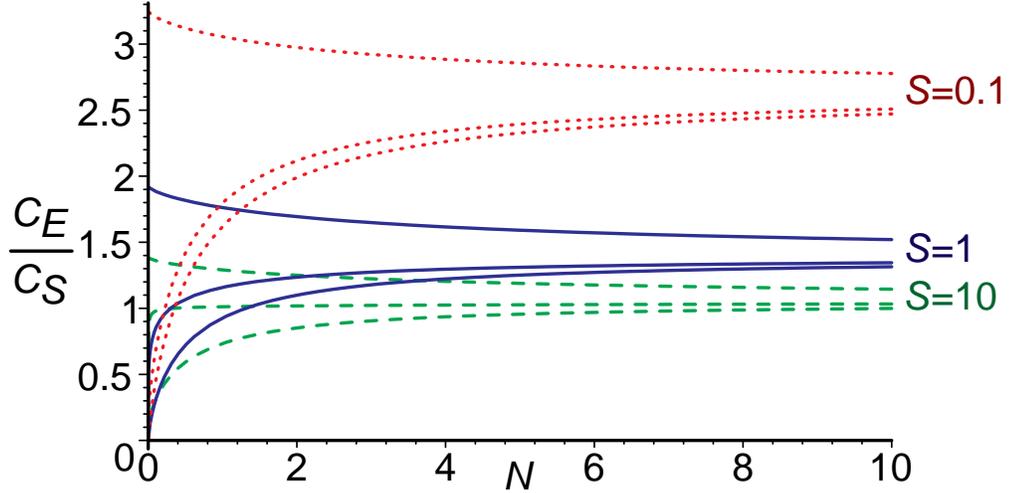}\medskip
\caption{This figure shows the curves given by the ratio of capacities
$C_E/C_{\rm{Shan}}$ for the quantum Gaussian channel with noise $N$ and the
nine combinations of values: amplification/attenuation parameter $k=0.1$, 
$1$, or $3$; and signal strength $S=0.1$, $1$, or $10$.  The dotted 
curves have $S=0.1$; the solid curves have $S=1$; and the dashed curves 
have $S=10$.  Within each set, the curves have the values $k=0.1$, $k=1$, 
and $k=3$ from bottom to top.
}
\label{CE-over-CS-1}
\end{figure}

\begin{figure}[tbp]
\epsfxsize=5in\hspace*{.5in}
\epsfbox{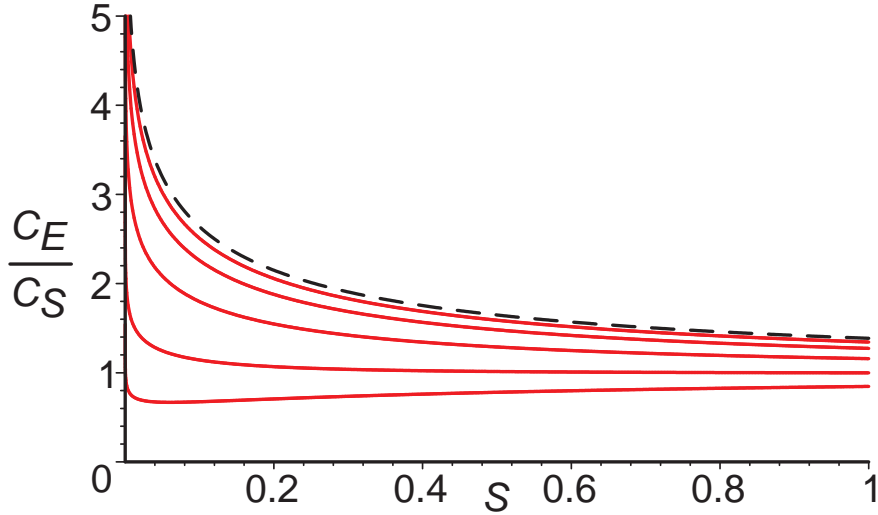}\medskip
\caption{The solid curves show the ratio of capacities 
$C_E/C_{\mathrm{Shan}}$ for the quantum Gaussian channel with signal 
strength $S$, amplification/attenuation paramter $k=1$ and noise 
$N=0.1$, $0.3$, $1$, $3$, and $10$ (from bottom to top).
The dashed curve is the limit of the solid curves as $N$ goes to $\infty$;
namely, $C_E/C_{\mathrm{Shan}} = (S+1)\log(1+1/S)$.  These curves approach
$\infty$ as $S$ goes to $0$, and approach $1$ as $S$ goes to $\infty$.
}
\label{CE-over-CS-2}
\end{figure}

The asymptotics of this formula are interesting.  Let us hold the signal
strength $S$ fixed, and let the noise $N$ go to infinity.  Then,
\beq
\lim_{N \rightarrow \infty} \frac{C_E}{C_{\rm Shan} } =
(S+1) \log \left(1 + \frac{1}{S}\right),
\eeq
which is independent of the attenuation/amplification parameter $k$.  This
ratio shows that the entanglement-assisted capacity can exceed the
Shannon formula by an arbitrarily large factor, albeit when the signal
strength $S$ is very small.  We have plotted $C_E/C_{\rm{Shan}}$ for
some parameters in Figs.\ \ref{CE-over-CS-1}~and~\ref{CE-over-CS-2}.

\begin{figure}[tbp]
\epsfxsize=5in\hspace*{.5in}
\epsfbox{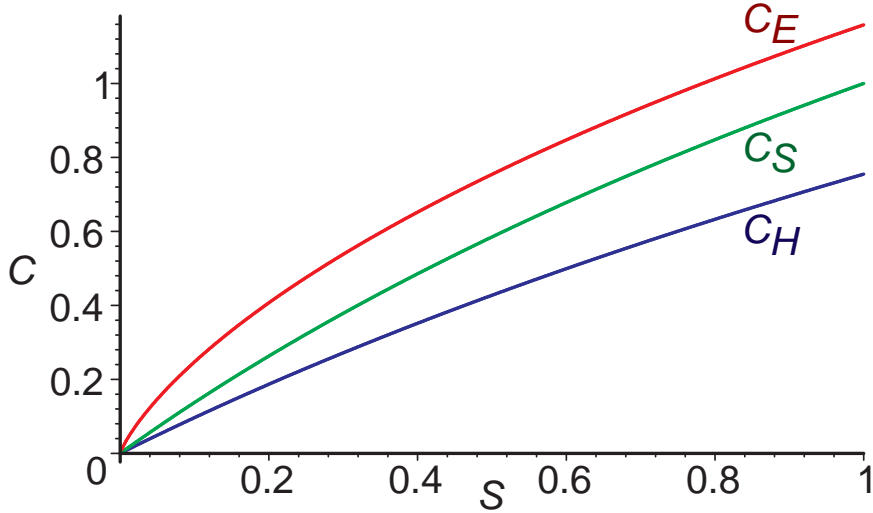}\medskip
\caption{The values of the capacities $C_E$, $C_{\mathrm{Shan}}$, and the 
conjectured $C_H$ (in units of bits) are plotted for the Gaussian channel 
with signal strength $S$, noise $N=1$,
and no amplification or attenuation ($k=1$).
As the curves approach $0$, their leading-order behavior is as follows:
$C_H \approx S$,
$C_{\mathrm{Shan}} \approx (\log_2 e) S$, and
$C_E \approx -\frac{1}{2} S\log_2 S$, so the ratios $C_E/C_{\mathrm{Shan}}$
and $C_E/C_H$ approach $\infty$ as $S$ goes to 0.
}
\label{three-curves}
\end{figure}

Possibly a better comparison than that of $C_E$ to $C_{\mathrm{Shan}}$
would be that of $C_E$ to $C_H$, as
$C_H$ is the best rate known for sending classical information over
a quantum channel without use of shared entanglement,
However, the optimal set of signal states to maximize $C_H$ for
Gaussian channels is not known.  For one-mode Gaussian channels with no
squeezing, it is conjectured to be a thermal distribution of coherent
states \cite{HW99}; if this conjecture is correct, then
$C_H \leq C_{\mathrm{Shan}}$ for these channels, so the ratio
$C_E/C_{\mathrm{Shan}}$ underestimates $C_E/C_H$;
see Fig.~\ref{three-curves}.

Some simple bounds on $C_E$ for the quantum Gaussian channel can be
obtained using the techniques of \cite{CEprl}.
Suppose that Alice takes a complex number $\alpha$, encodes it as
the state $\ket{\alpha}$, and sends this through a quantum Gaussian channel.
Bob then measures it in the coherent state basis.
Here, the measurement step adds 1 to the noise, and this channel
is thus equivalent to a classical Gaussian channel with average received
signal strength $k^2 S$, and average noise $N+1$ if $k \leq 1$,
$N+k^2$ if $k \geq 1$.
The quantum Gaussian channel must then have capacity greater than
the capacity of this classical Gaussian channel.
Conversely, Alice and Bob can simulate a quantum Gaussian channel by using
a classical complex Gaussian channel: Alice measures her state (in the
coherent state basis), sends the result through the classical channel,
and Bob prepares a coherent state that depends on the signal he receives.
If Alice starts with a state
$\ket{\alpha}$, when she measures it, she obtains a complex
number $\alpha + \epsilon$ where $\epsilon$ is a Gaussian with mean 0 and
variance $1$.  She can then multiply by $k^2$ to get
$k^2 \alpha + k^2 \epsilon$.  To simulate the quantum Gaussian
channel, she must send this state through a classical channel with noise
$N-k^2$ if $k \leq 1$, and $N-1$ if $k \geq 1$.  This classical channel must
then have classical capacity greater than $C_E$ for the quantum Gaussian
channel it is simulating.  The arguments in this
paragraph thus give bounds of
\beq
\log \left( 1 + \frac{k^2 S}{N+1} \right)  \leq C_E \leq
\log \left( 1 + \frac{S+1}{N/k^2-1} \right)
\label{r-0-bounds-littlek}
\eeq
for $k \geq 1$, and of
\beq
\log \left( 1 + \frac{S}{N/k^2+1} \right)  \leq C_E \leq
\log \left( 1 + \frac{k^2 (S+1)}{N-1} \right)
\label{r-0-bounds-bigk}
\eeq
for $k \leq 1$.  If we hold $S/N$ fixed, and let both these variables
go to infinity, we find that these bounds all go to $\log(1 + k^2S/N)$,
which corresponds to the classical Shannon bound (since the signal 
strength at the receiver is $k^2 S$). 

If $k=1$, we can compute better bounds than these based on continuous-variable
quantum teleportation and superdense coding.
Alice and Bob can use a shared entangled squeezed state to
teleport a continuous quantum variable \cite{cont-tel},
and can also use such a
state for a superdense coding protocol involving one channel use per shared
state that increases the classical
capacity of a quantum channel \cite{cont-dense}.
The squeezed state used, with squeezing parameter $r \geq 0$, is expressed in
the number basis as
\beq
\ket{s_r} =
\frac{1}{\cosh r} \sum_{j=0}^\infty (\tanh r)^j \ket{n_A=j}\ket{n_B=j},
\eeq
where $n_A$ and $n_B$ are the photon numbers in Alice's and Bob's modes,
respectively.
This state is squeezed, which means that it cannot be represented as a
mixture of coherent states with positive coefficients.
In this state, the uncertainty in the difference of Alice and Bob's
position coordinates, $x_A - x_B$, is reduced, as is the uncertainty
in the sum of their momentum coordinates $p_A + p_B$.  The conjugate
variables, $x_A + x_B$ and $p_A - p_B$, have increased uncertainty.  If
Alice and Bob measure their position coordinates, the difference of these
coordinates is a Gaussian variable with mean $0$ and variance $e^{-2r}$,
while the sum is a Gaussian with mean $0$ and variance $e^{2r}$.
Similarly, if they measure their momentum coordinates, the sum
has variance $e^{-2r}$ while the difference has variance $e^{2r}$.
Further, if either Alice's or Bob's
state is considered separately, it is a thermal state with average energy
$\sinh^2 r$.


In continuous-variable teleportation \cite{cont-tel},
Alice holds a state $\ket{t}$ she
wishes to send to Bob, and one half of the shared state $\ket{s_r}$.
She measures the difference of position coordinates of these states,
$x_m = x_t-x_A$, and the
sum of momentum coordinates, $p_m = p_t + p_A$.  These are commuting
observables, and so can be simultaneously determined.
She sends these measurement outcomes to Bob,
who then displaces his half of the shared state using $D(x_m + i p_m)$.

Using continuous-variable teleportation, Alice can simulate a quantum
Gaussian
channel with $k=1$, average input energy $S$ and noise $N$ by sending
the value $x_m + i p_m$ over
a classical complex Gaussian
channel with average input energy $S+(\cosh r)^2$ and noise $N-e^{-2r}$.
This gives a bound equal to the classical capacity of this channel:
\beq
C_E \leq \log\left(1 + \frac{S + (\cosh r)^2}{N-e^{-2r}}\right).
\label{r-bound-upper}
\eeq
Finding the $r$ which minimizes this expression gives
\beq
e^{2r} = \frac{D_1+1}{N}
\eeq
where
\beq
D_1 = \sqrt{(N+1)^2 + 4NS}
\eeq
is the value of the variable $D$ defined in Eq.~(\ref{general-D}) when
we set $k=1$.
This gives the bound
\beq
C_E \leq \log\left( 1 + \frac{S + (D_1 + N+1)/(2N)}{N}\right).
\eeq

Similarly, if Alice uses superdense coding \cite{cont-dense}
to send a continuous variable
to Bob, her protocol simulates a classical Gaussian channel.  The
average energy
input to this channel is $S-\sinh^2 r$ and the noise is $N+e^{-2r}$,
so we obtain the bound
\beq
C_E \geq \log \left(1 + \frac{S-\sinh^2 r}{N+e^{-2r}}\right).
\label{r-bound-lower}
\eeq
Maximizing this expression, we find the maximum is at
$e^{2r} = (D_1-1)/N$, and the bound obtained is
\beq
C_E \geq \log\left( 1 + \frac{S - (D_1 - N-1)/(2N)}{N}\right).
\eeq
Note that the bounds (\ref{r-bound-upper}) and (\ref{r-bound-lower})
reduce to the bounds of (\ref{r-0-bounds-littlek}) and
(\ref{r-0-bounds-bigk})
when there is no entanglement
in the squeezed state, i.e., when $r=0$.

\subsection{The Amplitude Damping Channel}

The amplitude damping channel describes a qubit channel which sends states
which decay by attenuation from $\ket{1}$ to $\ket{0}$, but which do not
undergo any other noise.  This channel can be described by two Krauss
operators,
\[
A_1 = \left(
\begin{array}{cc}
1 & 0 \\
0 & \sqrt{1-p}\\
\end{array}
\right)
\rm{\ \ \ and \ \ \ }
A_2 = \left(
\begin{array}{cc}
0 & \sqrt{p}\\
0 & 0 \\
\end{array}
\right)
\]
where
\[
{\cal N}: \rho \rightarrow \sum_{j=1}^2 A_j \rho A_j^\dag \,.
\]

\begin{figure}[tbp]
\epsfxsize=5.65in
\epsfbox{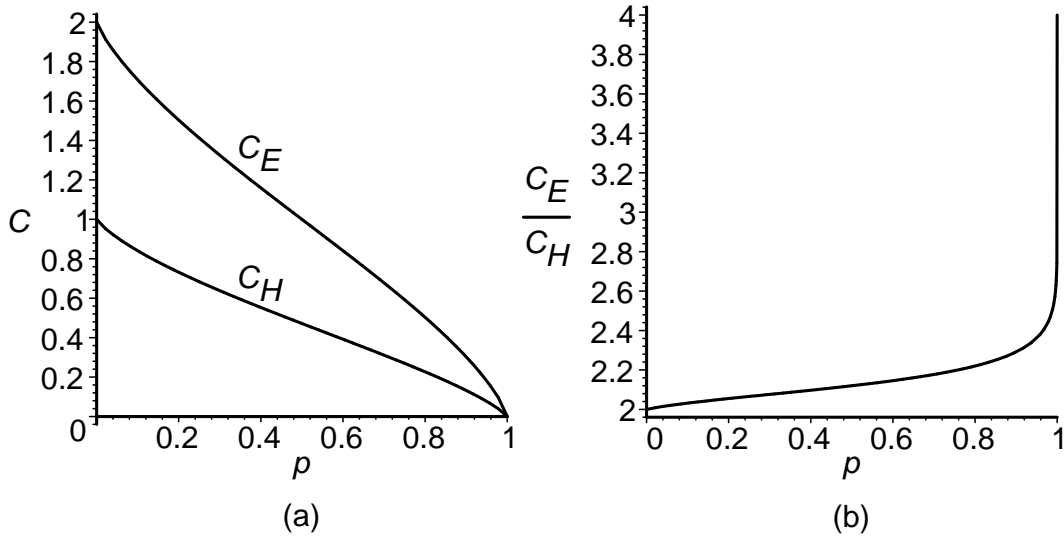}\medskip
\caption{(a) The capacity functions $C_E$ and $C_H$ for the amplitude 
damping channel are plotted against the damping
probability $p$.  (b) The ratio $C_E/C_H$ is plotted.
This curve is so steep near $p=1$ that for $p=1\!  - \! 10^{-50}$, the
computed value of the ratio
$C_E/C_H$ was only 3.8; the limiting value of 4 for $p=1$ was derived
analytically.
}
\label{damplitude-plots}
\end{figure}

The maximization over $\rho$ to find $C_E$ can be reduced to an
optimization over one parameter, as symmetry considerations show
that $\rho$ is of the form
\[
\rho_x = \left(\begin{array}{cc} 1-x & 0 \\ 0 & x \end{array}\right),
\]
This makes the optimization numerically tractable, and
the dependence of $C_E$ on $p$ is shown in
Fig.~\ref{damplitude-plots}.
As the damping probability $p$ goes to one, we can analytically
find the highest-order term in the expression for $C_E$, giving
\beq
C_E \approx - x (1-p)\log(1-p)
\eeq
for $0 < x < 1$.
Here we use ``$\approx$'' to mean that the ratio of the two sides
approaches 1 as $p$ goes to 1.

For the same channel, $C_H$ can also be obtained by optimizing
over a one-parameter family which uses two signal states $\rho_{x,+}$
and $\rho_{x,-}$
with equal probability\cite{Fuchs}.  These signal states are
\beq
\rho_{x,\pm} = \left(\begin{array}{cc} 1-x & \pm\sqrt{x(1-x)} \\
\pm\sqrt{x(1-x)} & x \end{array}\right).
\eeq
As $p$ goes to one, again we can analytically find the highest-order
term for $C_H$, which is
\beq
C_H  \approx - x(1-x)(1-p)\log(1-p).
\eeq
Thus, as $p$ goes to 1, the values of $x$ maximizing $C_E$ and $C_H$
respectively approach $1$ and $1/2$, and
the ratio $C_E /C_H $ approaches four.  These functions are shown
graphically in Fig.~\ref{damplitude-plots}.  In our previous paper
\cite{CEprl}, we showed that for the qubit depolarizing channel,
the ratio $C_E/C_H$
approached 3 as the depolarizing probability approached 1, and for
the $d$-dimensional depolarizing channel, the ratio approached $d+1$.
We do not know whether this ratio is bounded for finite-dimensional
channels, although we suspect it to be.  If so, then the
interesting question arises of how this bound depends on the dimensions
$\dim {\cal H}_{\rm{in}}$ and
$\dim {\cal H}_{\rm{out}}$\footnote{A. Holevo has found a qubit channel
where this ratio is 5.0798 \cite{newHolevo}}.

\section{Classical Reverse Shannon Theorem}
Shannon's celebrated noisy channel coding theorem established the
ability of noisy channels to simulate noiseless ones, and allowed
a noisy channel's capacity to be defined as the asymptotic
efficiency of this simulation. The reverse problem, of using a
noiseless channel to simulate a noisy one, has received far less
attention, perhaps because noisy channels are not thought to be a
useful resource in themselves (for the same reason, there has been
little interest in the reverse technology of water
desalination---efficiently making salty water from fresh water and
salt). We show, perhaps unsurprisingly, that any noisy discrete
memoryless channel of capacity $C$ can be asymptotically simulated
by $C$ bits of noiseless forward communication from sender to
receiver, given a source $R$ of random information shared
beforehand between sender and receiver. If this were not the case,
characterization of the asymptotic properties of classical
channels would require more than one parameter, because there
would be cases where two channels of equal capacity could not
simulate one another with unit asymptotic efficiency. In terms of
the desalination analogy, water from two different oceans might
produce equal yields of fresh drinking water, yet still not be
equivalent because they produced unequal yields of partly saline
water suitable, say, for car washing.

Although it is of some intrinsic interest as a result in classical
information theory, we view the classical reverse Shannon theorem
mainly as a heuristic aid in developing techniques that may
eventually establish its quantum analog, namely the conjectured
ability of all quantum channels of equal $C_E$ to simulate one
another with unit asymptotic efficiency in the presense of shared
entanglement.

Here we show that any classical discrete memoryless channel $N$,
of capacity $C$, can be asymptotically simulated  by $C$ uses of a
noiseless binary channel, together with a supply of prior random
information $R$ shared between sender and receiver.

The channel $N$ is defined by its stochastic transition matrix
$N_{yx}$ between inputs $x\in\{1...d_I\}$ and outputs
$y\in\{1...d_O\}$.  Let $N^n$ denote the extended channel
consisting of $n$ parallel applications of $T$, and mapping
$x\in\{1...d_I^n\}$ to $y\in\{1...d_O^n\}$.

\begin{theo}[Classical Reverse Shannon Theorem]\label{CRST}
Let $N$ be a DMC with Shannon capacity $C$ and $\epsilon$ a
positive constant.  Then for each block size $n$ there is a
deterministic simulation protocol $S_n$ for $N^n$ which makes use
of a noiseless forward classical channel and prior random
information (without loss of generality a Bernoulli sequence $R$)
shared between sender and receiver. When $R$ is chosen randomly,
the number of bits of forward communication used by the protocol
$S_n$ on channel input $x\in\{1...d_I\}^n$ is a random variable;
let it be denoted $m_n(x)$. The simulation is exactly faithful in
the sense that for all $n$ the stochastic matrix for $S_n$, when
$R$ is chosen randomly, is identical to that for $N^n$, \beq
\forall_{nxy} (S_n)_{yx}=(N^n)_{yx}, \eeq and it is asymptotically
efficient in the sense that the probability that the protocol uses
more than $n(C+\epsilon)$ bits of forward communication approaches
zero in the limit of large $n$, \beq
\lim_{n\rightarrow\infty}\max_{x\in\{1...d_I\}^n}
P(m_n(x)>n(C+\epsilon))=0. \eeq
\end{theo}

Note that the notion of simulation used here is stronger than the
conventional one used in the forward version of Shannon's noisy
channel coding theorem, and in eq. (\ref{MNsim}) defining the
generalized capacity of one quantum channel to simulate another.
There the simulations are required only to be asymptotically
faithful and their cost $m$ is deterministically upper bounded by
$n(C+\epsilon)$.  By contrast our simulations are exactly faithful
for all $n$ and their cost is upper bounded by $n(C+\epsilon)$
only with probability approaching 1 in the limit of large $n$, for
all $\epsilon>0$. To convert one of our simulations into a
standard one, it suffices to discontinue the simulation and
substitute an arbitrary output whenever $m_n(x)$ is about to
exceed $n(C+\epsilon)$.

To illustrate the central idea of the simulation, we prove the
theorem first for a binary symmetric channel (BSC), then extend
the proof to a general discrete memoryless channel.  Let $N$ be a
binary symmetric channel of crossover probability $p$.  Its
capacity $C$ is $1\!-\!H_2(p)=1+p\log_2
p+(1\!-\!p)\log_2(1\!-\!p)$. To prove the theorem in this case it
suffices to show that for any rate $\epsilon>0$, there is a
sequence of simulation protocols $S_n$ such that \beq
\forall_{nxy} (S_n)_{yx}=(N^n)_{yx},\label{BSC1} \eeq and \beq
\lim_{n\rightarrow\infty}\max_{x\in\{1...d_I\}^n}
P(m_n(x)>n(C+\epsilon))=0. \label{BSC2} \eeq

The simulation protocol $S_n$ is as follows:
\begin{enumerate}
\item  Before receiving the input $x\in\{0,1\}^n$, Alice and Bob use the random
information $R$ to choose a random set $Z(R,n)$ of
$2^{n(C+\epsilon/2)}$ $n$-bit strings. [We use $\epsilon/2$,
rather than $\epsilon$, to keep the total overhead, including
other costs, below $\epsilon$].
\item Alice receives the $n$-bit input $x$.
\item Alice simulates the true channel $N^n$ within her laboratory, obtaining an
$n$-bit ``provisional output'' $y$.  Although this $y$ is
distributed with the correct probability for the channel output,
she tries to avoid transmitting $y$ to Bob, because doing so would
require $n$ bits of forward communication, and she wishes to
simulate the channel accurately while using less forward
communication.  Instead, where possible, she substitutes a member
of the preagreed set $Z(R,n)$, as we shall now describe.
\item Alice computes the Hamming distance, $d=|x\!-\!y|$ between $x$ and $y$.
\item Alice determines whether there are any strings in the preagreed set $Z(R,n)$ having
the same Hamming distance $d$ from $x$ as $y$ does.  If so, she
selects a random one of them, call it $y'$, and sends Bob $0i$,
where $i$ is the approximately $n(C+\epsilon/2)$-bit index of $y'$
within the set $Z(R,n)$.  If not, she sends Bob the string $1y$,
the original unmodified $n$-bit string $y$, prefixed by a 1.
\item Bob emits $y'$ or $y$, whichever he has received, as the final output of the
simulation.
\end{enumerate}

It can readily be seen that the probability of failure in step
5---i.e., of there being no string of the correct Hamming distance
in the preagreed set $Z(R,n)$---decreases exponentially with $n$
as long as $\epsilon>0$. Thus the probability of needing to use
more than $C(1+\epsilon)$ bits of forward communication
approaches zero as required by Eq.~(\ref{BSC2}). On the other
hand, regardless of whether step 5 succeeds or fails, the final
output is correctly distributed (satisfying Eq.~(\ref{BSC2}))
since it has the correct distribution of Hamming distances from
the input $x$, and, for each Hamming distance, is equidistributed
among all strings at that Hamming distance from $x$. The theorem
follows.

For a general discrete memoryless channel the protocol must be
modified to take account of the nonbinary input and output
alphabets, and the fact that the output entropy may be different
for different inputs, unlike the BSC case. The notion of Hamming
distance also needs to be generalized. The new protocol uses the
notion of {\em type class\/}\cite{CT,C-K}. Two $n$-character
strings belong to the same type class if they have equal letter
frequencies (for example four a's, three b's, twelve c's etc.),
and are therefore equivalent under some permutation of letter
positions. We will consider input type classes (ITCs) and joint
input/output type classes (JTC), the latter being defined as a
set of input/output pairs (x,y) equivalent under some common
permutation of the input and output letter positions. In other
words, $(x_1,y_1)$ and $(x_2,y_2)$ belong to the same JTC if and
only if there exists a permutation of letter positions, $\pi$,
such that $\pi(x_1)=x_2$ and $\pi(y_1)=y_2$. Evidently, for any
given input and output alphabet size, the number of ITCs, and the
number of JTC are each polynomial in $n$. Let $k=1,2...K_n$
index the ITCs, and $\ell=1,2...L_n$ the JTC for inputs of
length $n$. The JTC will be our generalization of the Hamming
distance, since the transition probability $(N^n)_{yx}$ is equal
for all pairs $(x,y)$ in a given JTC. The new protocol follows:

\begin{enumerate}
\item Before receiving the input $x\in\{1,d_I^n\}$, Alice and Bob
use the common random information $R$ to preagree on $K_n$ random
sets $\{Z(R,n,k):k=1...K_n\}$ of $n$-letter output strings, one
for each ITC.  The set $Z(R,n,k)$ has cardinality
$2^{n(C_k+\epsilon/2)}$, where $C_k<C$ is the channel's capacity
for inputs in the $k$'th ITC (in other words, $1/n$ times the
channel's input:output mutual information on $n$-letter inputs
uniformly distributed over the $k$'th  ITC).  In contrast to the
BSC case, where the members of $Z(R,n)$ were chosen randomly from
a uniform distribution on the output space, the elements of
$Z(R,n,k)$ are chosen randomly from the (in general nonuniform)
output distribution induced by a uniform distribution of channel
inputs over the $k$'th ITC.
\item Alice receives the $n$-letter input $x$, determines which ITC,
$k$, it belongs to, and sends $k$ to Bob, using $o(n)$ bits to do
so.
\item Alice simulates the true channel $N^n$ in her laboratory,
obtaining an $n$-letter provisional output string $y$. Although
this $y$ is distributed with the correct probability for the
channel input $x$, she tries to avoid transmitting $y$ to Bob,
because to do so would require too much forward communication.
Instead she proceeds as described below.
\item Alice computes the index $\ell$ of the JTC to which the input/output pair
$(x,y)$ belongs.  As noted above, this JTC index is the
generalization of the Hamming distance, which we used in the BSC
case.
\item Alice determines whether there are any output strings in the
preagreed set $Z(R,n,k)$ having the same JTC index relative to
$x$ as $y$ does. If so, she selects a random one of them, call it
$y'$, and sends Bob the string $0i$ where $i$ is the approximately
$n(C+\epsilon/2)$-bit index of $y'$ within the set $Z(R,n,k)$. If
not, she sends Bob the string $1y$.
\item Bob emits $y'$ or $y$, whichever he has received, as the final output
of the simulation.
\end{enumerate}

This protocol deals with the problem of dependence of output
entropy on input by encoding each ITC separately. Within any one
ITC, the output entropy is independent of the input. The
communication cost of telling Bob in which ITC the input lies is
polylogarithmic in $n$, and so asymptotically negligible compared
to $n$. Because one cannot increase the capacity of a channel by
restricting its input, $nC$ is an upper bound the input:output
mutual information $nC_k$ for inputs restricted to a particular
ITC. Moreover, for any ITC $k$ and any input $x$ in that ITC, the
input:output pairs generated by the true channel $T^n$, will be
narrowly concentrated, for large $n$, on JTC whose transition
frequencies approximate (to within $O(\sqrt{n}))$ their asymptotic
values. Therefore, as before, for any $\epsilon>0$, the
probability of failure in step 5 will decrease exponentially with
$n$.  And as before, the simulated transition probability
$(S_n)_{yx}$ on each ITC is exactly correct even for finite $n$.
The reverse Shannon theorem for a general DMC follows, as does the
following corollary.

\begin{coro}[Efficient simulation of one noisy channel by another]
\hspace*{.1in} \\ In the presence of shared random information between sender and
receiver, any two classical channels of equal capacity can
simulate one another, in the sense of eq.~(\ref{MNsim}), with unit
asymptotic efficiency.
\end{coro}

 From the proof of the main theorem it can also be seen that when
inputs to the noisy channel being simulated come from a source
having a frequency distribution $q$ differing from the optimal one
$p$ for which capacity $C$ is attained, then the asymptotic cost
of simulating the channel on that source is correspondingly less.

\begin{coro}[Efficient simulation of noisy channels on constrained
sources] Let $N$ be a DMC, $q$ be a probability distribution over
the source alphabet, and $I(N,q)$ be the channel's constrained
capacity, equal to the single-letter input:output mutual
information on source $q$. Then, in the presence of shared random
information $R$ between sender and receiver, the action of $N$ on
any extended source having $q$ for each of its marginal
distributions can be simulated in the manner of Theorem \ref{CRST}
with perfect fidelity and a forward noiseless communication cost
asymptotically approaching $I(N,q)$: viz.
$\forall_\epsilon\lim_{n\rightarrow\infty}
P(m_n>n(I(N,q)+\epsilon))=0$.  Here $m_n$ denotes the number of
bits of forward communication used by the protocol when $R$ is
chosen randomly with a uniform distribution and inputs are chosen
randomly according to the constrained extended source.
\end{coro}

\section{Discussion---Quantum Reverse Shannon Conjecture}
We conjecture (QRSC) that in the presence of unlimited shared
entanglement between sender and receiver, all quantum channels of
equal $C_E$ can simulate one another with unit asymptotic
efficiency, in the sense of eq.~(\ref{MNsim}). By the results of
the previous section, the conjecture holds for classical channels
(where the shared random information required for the classical
reverse Shannon theorem is obtained from shared entanglement). In
our previous paper \cite{CEprl} we showed that the QRSC also holds
for another class of channels, the so-called Bell-diagonal
channels, which commute with teleportation and superdense coding.
For these channels, the single-use entanglement-assisted classical
capacity of the channel via superdense coding is equal to the
forward classical communication cost of simulating it via
teleportation. The QRSC asserts this equality holds asymptotically
for all quantum channels, even when (as for the amplitude damping
channel) it is does not hold for single uses of the channel. We
hope that the arguments used to prove the classical reverse
Shannon theorem can be extended to demonstrate its quantum analog.

If the QRSC is true, one useful corollary would be the inability
of a classical feedback channel from Bob to Alice to increase $C_E$. A
causality argument shows that a feedback channel cannot increase $C_E$
for noiseless quantum channels. If we could simulate noisy quantum
channels by noiseless ones, this would imply that if a feedback
channel increased $C_E$ for any noisy channel, it would have to
increase $C_E$ for noiseless ones as well, violating causality.

We thank Igor Devetak, David DiVincenzo, Alexander Holevo, Michael
Nielsen and Barbara Terhal for helpful discussions, and the
referees for careful reading and advice resulting in significant
improvements.


\begin{thebibliography}{10}
\setlength{\itemsep}{0pt}
\setlength{\parsep}{0pt}
\setlength{\parskip}{0pt}

\bibitem{AS} N. Alon and J. H. Spencer, {\em The Probabilistic Method,}
John Wiley and Sons, New York (1991).
\bibitem{AK} A. Ashikhmin and E. Knill, ``Nonbinary quantum stabilizer codes,''
{\em IEEE Trans.\ Inf.\ Theory} {\bf 47,} pp. 3065--3072 (2001);
LANL eprint quant-ph/0005008.
\bibitem{BNS97}H. Barnum, M.~A. Nielsen, and B. Schumacher, ``Information
transmission through noisy quantum channels,'' {\em Phys.\  Rev.\  A} {\bf 57,}
pp. 4153--4175 (1998);  LANL eprint quant-ph/9702049.
\bibitem{BST}H. Barnum, J. A. Smolin, and B. M. Terhal, ``The quantum
capacity is properly defined without encodings,''
{\em Phys.\ Rev.\  A} {\bf 58,} pp.\ 3496--3501 (1998); LANL e-print
quant-ph/9711032.
\bibitem{BBPS} C.~H. Bennett, H.~J. Bernstein, S. Popescu, and B. Schumacher,
``Concentrating partial entanglement by local operations,''
{\em Phys.\  Rev.\  A\/} {\bf 53,} pp.\ 2046--2052 (1996).
\bibitem{teleport} C.~H. Bennett, G. Brassard, C. Cr\'epeau, R. Jozsa,
A Peres and W.~K. Wootters,
``Teleporting an unknown quantum state via dual classical and EPR channels,''
{\em Phys.\  Rev.\  Lett.\/} {\bf 70,} pp.\ 1895--1899 (1993).
\bibitem{CEprl}C.~H. Bennett, P.~W. Shor, J.~A. Smolin, and A.~V. Thapliyal,
``Entanglement-assisted capacity of noisy quantum channels,''
{\em Phys.\ Rev.\ Lett.\/} {\bf 83}, pp.\ 3081--3084 (1999); LANL eprint
quant-ph/9904023.
\bibitem{Bennett-Shor} C. H. Bennett and P. W. Shor, ``Quantum information
theory,'' {\em IEEE Trans.\ Inform.\ Theory} {\bf 44}, pp. 2724--2742
(1998).
\bibitem{superdense}  C. H. Bennett and S. J. Wiesner, ``Communication via
one- and two-particle operators on Einstein-Podolsky-Rosen states,''
{\em Phys. Rev. Lett.\/} {\bf 69}, pp. 2881--2884 (1992).
\bibitem{cont-tel} S.~L. Braunstein and H.~J. Kimble, ``Teleportation of
continuous quantum variables,'' {\em Phys.\ Rev.\ Lett.\/}
{\bf 80}, pp.\ 869--872 (1998).
\bibitem{cont-dense} S. L. Braunstein and H. J. Kimble, ``Dense coding
with continuous quantum variables,'' {\em Phys.\ Rev.\ A} {\bf 61,}
art.\ 042302 (2000).
\bibitem{CandA} N. Cerf and G. Adami, ``Von Neumann capacity
of noisy quantum channels, {\em Phys.\ Rev.\ A\/} {\bf 56,}
pp.\ 3470--3483 (1997).
\bibitem{CT} Cover and Thomas, {\em Elements of Information Theory,}
John Wiley and Sons, New York (1991).
\bibitem{C-K} I. Csisz\'ar and J. K\"orner, {\em Information Theory:
Coding Theorems for Discrete Memoryless Systems,} Akad\'emiai Kiad\'o,
Budapest (1981).
\bibitem{boundecUS} D. P. DiVincenzo, T. Mor, P. W. Shor, J. A. Smolin
and B. M. Terhal, ``Unextendible product bases, uncompletable product bases,
and bound entanglement,''
{\em Phys. Rev. Lett.\/} {\bf 82}, pp.\ 5385-5388, (1999);
LANL eprint quant-ph 9908070.
\bibitem{Fuchs} C. A. Fuchs, personal communication.
\bibitem{Haus} P. Hausladen, R. Jozsa, B. Schumacher, M. Westmoreland, and
W. K. Wootters, ``Classical information capacity of a quantum channel,''
{\em Phys.\ Rev.\ A} {\bf 54,} pp.\ 1869--1876 (1996).
\bibitem{HW99} A.~S. Holevo and R.~F. Werner ``Evaluating capacities of
bosonic Gaussian channels,'' {\em Phys.\ Rev.\ A} {\bf 63}, art.\ 032313
(2001); LANL eprint quant-ph/9912067.
\bibitem{Holevo} A. S. Holevo, ``The capacity of the quantum channel with
general signal states,'' {\em IEEE Trans. Information Theory} {\bf 44,}
pp.~269--273 (1998).
\bibitem{newHolevo} A. S. Holevo, ``On entanglement-assisted classical
capacity,'' LANL eprint quant-ph/0106075.
\bibitem{boundecH} P. Horodecki, M. Horodecki, and R. Horodecki, ``Binding
entanglement channels,'' {\em J. Modern Optics} {\bf 47,} pp.\ 347--354
(2000), LANL eprint quant-ph/9904092.
\bibitem{HJW}  L. P. Hughston, R. Jozsa, W.~K. Wootters,
``A complete classification of quantum
ensembles having a given density matrix,'' {\em Phys. Lett. A} {\bf 183},
pp.~14-18 (1993).
\bibitem{JS} R. Jozsa and B. Schumacher, ``A new proof of the
quantum noiseless coding theorem,'' {\em J. Modern Optics}
{\bf 41}, pp.\ 2343--2350 (1994).
\bibitem{CKing} C. King, ``Additivity for unital qubit channels,''
{\em J. Math.\ Phys.\/}, to appear; LANL eprint quant-ph/0103156.
\bibitem{LR} E. Lieb and M.B. Ruskai,
``Proof of the Strong Subadditivity of Quantum Mechanical Entropy''
{\em J. Math.\ Phys.\/}  {\bf 14}, pp.~1938--1941  (1973).
\bibitem{Lindblad91} G. Lindblad, ``Quantum entropy and quantum
measurements,'' in {\em Quantum aspects of optical communications,
Lecture Notes in Physics 378}, C. Bendjaballah, O. Hirota, S. Reynaud (eds.)
Springer, pp.\ 71--80 (1991).
\bibitem{LP99}H.-K. Lo and S. Popescu,
``The classical communication cost of entanglement manipulation:
Is entanglement an inter-convertible resource?''
{\em Phys.\ Rev.\ Lett.\/} {\bf 83,}
pp.\ 1459--1462 (1999); LANL eprint quant-ph/9902045.
\bibitem{NC} M. A. Nielsen and I. L. Chuang, {\em Quantum Computation
and Quantum Information}, Cambridge University Press, Cambridge, UK, 2000.
\bibitem{S} B. Schumacher, ``Quantum coding,'' {\em Phys.\  Rev.\  A}
{\bf 51}, pp.\ 2738--2747 (1995).
\bibitem{SN}
B. W. Schumacher and M. A. Nielsen, ``Quantum data processing and error
correction'', {\em Phys.\  Rev.\ A} {\bf 54,} pp.~2629--2635, (1996).
\bibitem{SW}
B. Schumacher and M. D. Westmoreland, "Sending classical information via
noisy quantum channels," {\em Phys. Rev. A} {\bf 56},
pp.\ 131--138 (1997).
\bibitem{VC} G. Vidal and J. I. Cirac, ``Irreversibility in asymptotic
manipulations of entanglement,'' {\em Phys.\ Rev.\ Lett.\/} {\bf 86},
pp.~5803-5806 (2001); LANL eprint quant-ph/0102036.
\bibitem{wal-mil} D.~F. Walls and G.~J. Milburn,
{\em Quantum Optics}, Springer Verlag, Berlin (1994).
\bibitem{Wehrl} A. Wehrl, ``General properties of entropy,''
{\em Rev.\ Mod.\ Phys.\/} {\bf 40} pp.~221--260 (1978).
\end{thebibliography}
\end{document}